\documentclass[fleqn,usenatbib]{mnras-arxiv}

\usepackage[T1]{fontenc}
\usepackage{ae,aecompl}

\usepackage{graphicx}	
\usepackage{amsmath}	
\usepackage{amssymb}	
\usepackage{mathtools}
\usepackage{multicol}
\usepackage[flushleft]{threeparttable}
\usepackage[utf8]{inputenc}
\usepackage{multirow}
\usepackage{booktabs}
\usepackage{siunitx}
\usepackage{mathptmx}
\usepackage{balance}
\usepackage[colorinlistoftodos,textsize=small,textwidth=12mm]{todonotes}
\usepackage{marginnote}
\usepackage{enumerate}
\usepackage{framed}
\usepackage{soul,color}
\soulregister\cite7
\soulregister\ref7
\soulregister\pageref7

\newcommand{\notel}[2][]{{%
 \let\marginpar\marginnote
 \reversemarginpar
 \todo[#1]{#2}}}

\newcommand{\noteAndre}[1]{
\let\marginpar\marginnote
\reversemarginpar
\todo[color=green!60]{#1}}

\newcommand{\noteGPC}[1]{
\let\marginpar\marginnote
\reversemarginpar
\todo[color=cyan!60]{#1}}

\colorlet{shadecolor}{yellow}

\title[A Framework for Designing and Evaluating Solar Flare Forecasting Systems]{A Framework for Designing and Evaluating Solar Flare Forecasting Systems}

\author[T. Cinto et al.]{
T. Cinto,$^{1,2}$\thanks{E-mail: tiago.cinto@pos.ft.unicamp.br}
A. L. S. Gradvohl,$^{1}$
G. P. Coelho$^{1}$
and A. E. A. da Silva$^{1}$
\\
$^{1}$School of Technology (FT), University of Campinas (UNICAMP), Paschoal Marmo st., 1888, Limeira - SP, 13484-332, Brazil\\
$^{2}$ Federal Institute of Education, Science and Technology of Rio Grande do Sul (IFRS) - Campus Feliz, Princesa Isabel st., 60, Feliz - RS, 95770-000, Brazil \\
}

\pubyear{2019}

\begin{document}
\label{firstpage}
\pagerange{\pageref{firstpage}--\pageref{lastpage}}
\maketitle

\begin{abstract}
Disturbances in space weather can negatively affect several fields, including aviation and aerospace, satellites, oil and gas industries, and electrical systems, leading to economic and commercial losses. Solar flares are the most significant events that can affect the Earth's atmosphere, thus leading researchers to drive efforts on their forecasting. The related literature is comprehensive and holds several systems proposed for flare forecasting. However, most techniques are tailor-made and designed for specific purposes, not allowing researchers to customize them in case of changes in data input or in the prediction algorithm. This paper proposes a framework to design, train, and evaluate flare prediction systems which
present promising results. Our proposed framework involves model and feature selection, randomized hyper-parameters optimization, data resampling, and evaluation under operational settings. Compared to baseline predictions, our framework generated some proof-of-concept models with positive recalls between 0.70 and 0.75 for forecasting $\geq M$ class flares up to 96 hours ahead while keeping the area under the ROC curve score at high levels.
\end{abstract}

\begin{keywords}
Sun: flares -- sunspots -- methods: data analysis -- Sun: activity -- Sun: X-rays, gamma rays -- techniques: miscellaneous
\end{keywords}



\section{Introduction}
The Sun has a very active atmosphere, featuring several events that directly impact all bodies in the solar system. Affected aspects include the solar wind, the near-Earth space, and the Earth's atmosphere. Those events can damage several fields, including aviation and aerospace, satellites, oil and gas industries, and electrical systems, leading to economic and commercial losses \citep{NRC:2009}. 
Solar flares are one of the most significant events since they comprehend sudden releases of radiation and particles that can affect the Earth's atmosphere in a few hours or minutes. Mostly related to active regions (ARs) \citep{Canfield:2001}, solar flares are releases of x-rays of 1 to 8 \AA{ngstr{\"o}m} (\si{\angstrom}) measured in watts per square meter (\si{\watt\per{\square\meter}}) \citep{Messerotti:2009}. Such events are classified in a scale ranging from A, B, C, M, and X, in which each class has a peak flux ten times higher than its predecessor (Table~\ref{tbl:classificacao-explosoes-solares}). 

Each class of flare also varies in a linear scale from 1 to 9, which represents the flare intensity. Then, flares are described by the product of its intensity factor with the x-ray peak value of its class.






\begin{table}
\centering
\caption{Solar flares classes.}
\label{tbl:classificacao-explosoes-solares}
\begin{tabular}{cc}
\toprule
\multirow{2}{*}{\textit{Class}} & \textit{Peak Flux in \si{\watt\per{\square\meter}}}\tabularnewline
                                & \textit{Between \SI{1}{\angstrom}~and~\SI{8}{\angstrom}}\tabularnewline
\midrule
A  & \num{<e-7} \tabularnewline

B  & \numrange{>=e-7}{<e-6} \tabularnewline

C  & \numrange{>=e-6}{<e-5} \tabularnewline

M  & \numrange{>=e-5}{<e-4} \tabularnewline

X  & \num{>=e-4} \tabularnewline
\bottomrule
\end{tabular}
\end{table}

Because of the several solar flares effects reported, it is critical to design systems to forecast such events. We are currently facing a hot topic that became recurrent in research agendas recently. 

\subsection{Flare Forecasting Efforts}

Many researchers propose the use of photospheric magnetic data to investigate ARs and their relationship with solar activity \citep{McAteer:2010}. Others, in turn, focus their studies on the investigation of ARs photospheric features \citep{McIntosh:1990} and magnetic topologies \citep{Hale:1919} concerning solar flares productivity. Regardless of the guiding principles, researchers have proposed a lot of methods to forecast the occurrence of solar flares. Here, the most notable examples include, but are not limited to: linear discriminant analysis \citep{Barnes:2007, Leka:2018}, Bayesian statistics \citep{Wheatland:2005, Yu:2010b, Zhang:2011}, neural networks \citep{Qahwaji:2006, Qahwaji:2007, Wang:2008, Colak:2009, Colak:2007, Ahmed:2013, Li:2013, Shin:2016, Hada-Muranushi:2016, Nishizuka:2018, Huang:2018, Park:2018, Domijan:2019}, decision trees \citep{Yu:2009, Yu:2010a, Zhang:2011, Huang:2010, Huang:2013}, radial basis functions \citep{Colak:2007, Qahwaji:2007, Qahwaji:2006}, learning vector quantization \citep{Yu:2009, Li:2013}, unsupervised learning vector quantization \citep{Li:2011}, Poisson statistics \citep{Gallagher:2002, Falconer:2011, Bloomfield:2012, Falconer:2014, McCloskey:2018}, support-vector machines \citep{Qahwaji:2007, Li:2008, Yuan:2010, Yang:2013, Bobra:2015, Muranushi:2015, Raboonik:2016, Nishizuka:2017, Sadykov:2017b, Domijan:2019}, superposed epoch analysis \citep{Mason:2010}, regression models \citep{Lee:2007, Song:2009, Yuan:2010, Muranushi:2015,Anastasiadis:2017}, AdaBboost \citep{Lan:2012}, random forest \citep{Liu:2017c, Domijan:2019}, image-case-based prediction \citep{Liu:2017b}, multi-model prediction \citep{Liu:2017a}, relevance-vector machine \citep{Al-Ghraibah:2015a}, the least absolute shrinkage and selection operator \citep{Benvenuto:2018, Jonas:2018}, multiple linear regression \citep{Shin:2016}, k-nearest neighbors \citep{Li:2008, Huang:2013, Winter:2015, Nishizuka:2017}, extremely randomized trees \citep{Nishizuka:2017}, unsupervised fuzzy clustering \citep{Benvenuto:2018}, linear classifiers \citep{Jonas:2018}, ensemble methods \citep{Huang:2010, Guerra:2015}, and expert systems \citep{Miller:1988, McIntosh:1990}.

Most of the aforementioned papers had something in common, the use of a machine learning or statistical technique to build their prediction models. Machine learning is a computer science branch aimed at learning from data and at making predictions on new observations, which is defined by the so-called classification supervised learning task \citep{Han:2006}. In classification, the user provides the learning algorithm with data examples (training samples) and their corresponding classes representing the existence of a particular event (in this case, flare or non-flare). 

Despite the high levels of performance and improvements achieved by these learning systems, such techniques are most of the times tailor-made and designed for specific purposes, not allowing researchers to flexibly customize them in case of data input changes or the need for new prediction algorithms, for instance. Thus, we propose a framework in an attempt to standardize the process of designing and evaluating solar flare predictors. Recently, few papers focused on such standardization to generate new forecasting models. To name some, we can cite \cite{Muranushi:2015} and \cite{Leka:2018}.

\cite{Leka:2018} proposed the Discriminant Analysis Flare Forecasting System (DAFFS). DAFFS evaluates the magnetic fields on the Sun for evidence of stored energy and for magnetic complexity known to be associated with flare productivity and  history. DAFFS uses near-real-time vector magnetic data along with reports from the National Oceanic and Atmospheric Administration (NOAA) Geostationary Operational Environmental Satellite (GOES). This tool comprehends an operational forecasting system that runs twice a day and make predictions just before 00:00 and 12:00 UT for several event definitions and validity periods (to name some, prediction of $\geq C1.0$, $\geq M1.0$, $\geq X1.0$ flares with cadences of 24 and 48 hours, and 24 hours of validity). 

DAFFS trains new and custom forecasting models on-demand considering as much data as possible, thus always using the period between 2012 and the most recent month. In addition, DAFFS provides some tweaks concerning the forecasting models trained, such as the definition of a custom flare magnitude threshold, the possibility of choosing the type of prediction (if it will be full-disk or AR-by-AR), and the adjustment of the precision vs. recall trade-off threshold. Although flexible, DAFFS presented some disadvantages such as being restricted to the discriminant analysis technique, it did not treat imbalanced data scenarios, and it did not adjust the complexity of models to avoid over-fitting. 

\cite{Muranushi:2015}, in turn, proposed the Universal Forecast Constructor by Optimized Regression of Inputs (UFCORIN), which was a generic time series predictor by definition. Their system could be set to predict any time series variable from an arbitrary set of input time series using linear regression. Thus, UFCORIN allowed users to flexibly change the input and the corresponding target parameters when more advanced data became available or when the need for building new models with different targets arose. 

However, the design of UFCORIN had some disadvantages, such as it only used linear regression as the prediction model. Basically, their pipeline could easily allow the change of time series inputs and targets outputs to generate new regression models flexibly. 

\subsection{Aims and Scope}

Here we propose a framework aiming to overcome the aforementioned restrictions and comprehend several advances in the proposal of creating some standardization when training prediction models for solar flares. Aspects like automated feature selection, hyper-parameters fine-tuning, imbalanced data resampling and spot-check of distinct models are some of the most notable contributions of our paper. Then, this research will present a framework created to design, train, and evaluate flare forecasting systems under operational settings with flexibility and performance. 

We divided this paper into six sections. In Section~\ref{sec:Dataset}, we detail the dataset prepared to evaluate the framework proposed as well as our flare catalog and parameters. In Section~\ref{sec:proposedFramework}, we explain the proposed framework, providing details of each step performed and techniques employed. In Section~\ref{sec:Results}, we underlie our results and show how the framework has improved the forecasting systems proposed. In Section~\ref{sec:comparisons-with-literature}, we will position our forecast performance within the literature. Finally, in Section 6, we underlie the conclusions of this study.

\section{Dataset}\label{sec:Dataset}
We collected data from the repositories maintained by the Space Weather Prediction Center (SWPC)\footnote{http://www.swpc.noaa.gov}. SWPC is one of the nine centers aimed at climate prediction in the US and is associated with NOAA, which focuses on oceanic and atmospheric conditions. It provides real-time monitoring of solar events that impact navigation, telecommunications, satellites,  and other systems. Data collected by NOAA/SWPC are freely available for study and research purposes.

We gathered and integrated data from two NOAA/SWPC's repositories\footnote{ftp://ftp.swpc.noaa.gov/pub/warehouse/}: Daily Solar Data (DSD), which observes the Sun's daily behaviors, and Sunspot Region Summary (SRS), including the sunspots magnetic classes recorded in DSD. 

The data referred to the period between January 01, 1997 and January 15, 2017. We obtained 7,320 records comprehending data from 17 different attributes, such as radio flux, sunspot number and area, x-ray flux, 3 identifiers representing the year, month and the day when data were collected, 3 features representing the amount of each flare class occurred, and also 7 binary attributes representing the existence in the photosphere of the most common magnetic classes of sunspots on a given day \citep{Hale:1919}. We explain these features as follows:

\begin{itemize}
\item \emph{Year}: self-explanatory, identifier attribute. 
\item \emph{Month}: self-explanatory, identifier attribute.
\item \emph{Day}: self-explanatory, identifier attribute.
\item \emph{X class flares}: number of X class flares occurred.
\item \emph{M class flares}: number of M class flares occurred.
\item \emph{C class flares}: number of C class flares occurred.

\item \emph{Radio flux}: the solar radio flux at \SI{10.7}{\cm} (\SI{2800}{\mega\hertz}) is a solar activity index. Also called the F10.7 index, it is one of the longest solar activity running records. Radio emissions originate high in the chromosphere and low in the solar atmosphere corona\footnote{\label{swpc-readme}ftp://ftp.swpc.noaa.gov/pub/indices/old\_indices/README}.

\item \emph{Sunspot area}: it refers to the sum of all observed sunspots areas and is measured in millionth units of the solar hemisphere\textsuperscript{\ref{swpc-readme}}.

\item \emph{X-ray flux}: it corresponds to the daily average background x-ray flux measured by the NOAA/SWPC primary GOES satellite. To calculate this value, sensors record 24 x-ray measures for a given day (one for each hour) and three groups with periods of eight hours are created with them. Then, the NOAA/SWPC records the lowest flux values of each group and calculates the average between the first and third group minimum values. Finally, they compare the resultant mean value with the second minimal value and report the lowest measurement as the x-ray background flux\textsuperscript{\ref{swpc-readme}}.

\item \emph{Sunspot number}: it refers to the number of sunspots computed on a given day. Also called Wolf's number of sunspots, it is given by $R = k (10g + s)$, where $k$ is a scalable factor representing the combined effects of observation conditions, $g$ is the number of observed ARs, and $s$ is the  number of sunspots inside the ARs\textsuperscript{\ref{swpc-readme}}. Here we clarify that the sunspot number is the one found within the DSD dataset (the SRS data also include the sunspot number of each NOAA numbered region; however, this is not the Wolf's sunspot number)\textsuperscript{\ref{swpc-readme}}.

\item \emph{Magnetic classes}: based on the Mt. Wilson's Taxonomy~\citep{Hale:1919}, these classes describe sunspots according to their magnetic complexity into some distinct categories. We encoded these categories into 7 binary attributes \citep{Sarkar:2018} representing the existence of the most common magnetic classes of sunspots on a given day. We decided to use only the most frequent classes based on the reports of \cite{Jaeggli:2016}, who analyzed the years between 1992 and 2015 and created a summary of magnetic classes occurrence that comprehended most of the data period we are using. We describe the magnetic classes considered as follows:
\begin{itemize}
\item \textit{Alpha}: a unipolar sunspot group;
\item \textit{Beta}: a bipolar sunspot group that has both positive and negative polarities with a simple division line between them;
\item \textit{Gamma}: a sunspot group with positive and negative polarities irregularly distributed (it is not possible to classify this group as bipolar);
\item \textit{Beta-gamma}: a bipolar spots group so complex that it is not possible to represent the separation line between the spots with opposite polarities;
\item \textit{Beta-delta}: a spots group with beta class and also with one or more delta class spots;
\item \textit{Beta-gamma-delta}: a sunspot group with beta-gamma class as well as with one or more delta class sunspots;
\item \textit{Gamma-delta}: a spots group with a gamma magnetic configuration, that has one (or more) delta sunspots.
\end{itemize}
\end{itemize}

\subsection{Data Preparation}
This section describes some aspects we had to deal when preparing the dataset to deliver more refined data to our framework. We needed to employ two distinct techniques for dataset preparation: missing data imputation and data standardization.

We chose to input missing data through a k-Nearest Neighbors model ($k$-NN) \citep{Han:2006}, a technique that inputs data considering the similarity between tuples (from now on, let a tuple be a dataset entry holding all the aforementioned features).  

By default, the $k$-NN algorithm uses the Euclidean distance as the proximity coefficient and can thus only be applied to numeric attributes. However, the assembled dataset held binary attributes along with the numeric ones. Thus, we coded the $k$-NN to use the Gower's distance \citep{Gower:1971}, a distance metric appropriate to a mixed-attribute scenario. 

The Gower's coefficient defines two distinct calculations, one for numeric attributes and one for the binary ones. The mean value of both calculations gives the distance score. We can calculate the nominal distance through Equation~\ref{eq:gower-atributos-nominais}.

\begin{equation}
\label{eq:gower-atributos-nominais}
\begin{gathered}
S_{ij}=\frac{1}{p}\sum_{i=1}^{p}S_{i}
\end{gathered}
\end{equation}

\noindent where $S_{ij}$ is the similarity between objects $i$ and $j$; $S_{i} = 1$ is attributed when data of objects $i$ and $j$ match; $S_{i} = 0$ is attributed when data of objects $i$ and $j$ are different; and $p$ is the number of variables. 

In turn, we calculate the distance between numeric attributes in Gower's schema through the normalized Manhattan distance, given in Equation~\ref{eq:gower-atributos-numericos}.

\begin{equation}
\label{eq:gower-atributos-numericos}
\begin{gathered}
S_{i}=1- \left | \frac{y_{ik} - y_{jk}}{\max(y_{k}) - \min(y_{k})} \right |
\end{gathered}
\end{equation}

\noindent where $y_{ik}$ is the object $i$ \emph{k-th} variable value; $y_{jk}$ is the object $j$ \emph{k-th} variable value; $\max(y_{k})$ is the \emph{k-th} variable maximum value; and $\min(y_{k})$ is the \emph{k-th} variable minimum value.

In addition to missing data imputation, the dataset also held some discrepancy issues in features data ranges. For instance, while the sunspot number and area, and radio flux data were ranging between \big[0,401\big], \big[0,5690\big], and \mbox{\big[65,298\big]}, respectively, x-ray flux data ranged between the interval \mbox{\big[$10^{-7}$, $2\times10^{-5}$\big]}. Those discrepancy issues are known to severely damage the performance of predictors, thus leading us to perform z-score normalization with the numeric attributes \citep{Han:2006}. 

Also known as the standard score, z-score is a technique that transforms data based on their interval mean and standard deviation. Some papers argue in favor of using the standard score instead of an ordinary min-max normalization since it positively affects the predictive performance \citep{Nishizuka:2017}. 



\subsection{Sliding time window}
Aware of the benefits of including the data evolution in time series, we also decided to re-organize our data in a more appropriate way. This section will describe some aspects of the sliding time window we designed.

Learning systems that must deal with data evolution over a period of time and make forecasts in a sequential and supervised fashioned way are called short-term predictors \citep{Yu:2009}. To design data for these classifiers, one should use the sliding time window schema, whose nature represents the data evolution $n$ days before some event occurrence. Several papers focus on proving the benefits of using this time window combined with flares forecasting models, such as \cite{Yu:2009}, \cite{Yu:2010b}, and \cite{Huang:2010}. 

The main principle of the sliding time window relies on the solar data observation at the $t$ instant (from now on, let the $t$ instant be any day) and some days before $t$, i.e., $[t - \Delta t]$. The interval between $t$ and $[t - \Delta t]$ is the sliding time window. 

A key issue of such schema is to choose a reasonable window length. If the window is too short, data represented may not be enough. On the other hand, useless data may be inputted in case of longer windows. Radio flux emissions last from 3 to 5 days, so it is reasonable to consider this period when adjusting the window length \citep{Yu:2009}. Hence, we designed our data stream 4 days before the $t$ instant:

\begin{itemize}
\item \textit{sunspost number} $\big[t-4d,t-3d,t-2d,t-1d,t\big]$;
\item \textit{sunspost area} $\big[t-4d,t-3d,t-2d,t-1d,t\big]$;
\item \textit{x-ray flux} $\big[t-4d,t-3d,t-2d,t-1d,t\big]$;
\item \textit{radio flux} $\big[t-4d,t-3d,t-2d,t-1d,t\big]$;
\item \textit{alpha class} $\big[t-4d,t-3d,t-2d,t-1d,t\big]$;
\item \textit{beta class} $\big[t-4d,t-3d,t-2d,t-1d,t\big]$;
\item \textit{gamma class} $\big[t-4d,t-3d,t-2d,t-1d,t\big]$;
\item \textit{beta-gamma class} $\big[t-4d,t-3d,t-2d,t-1d,t\big]$;
\item \textit{beta-delta class} $\big[t-4d,t-3d,t-2d,t-1d,t\big]$;
\item \textit{beta-gamma-delta class} $\big[t-4d,t-3d,t-2d,t-1d,t\big]$;
\item \textit{gamma-delta class} $\big[t-4d,t-3d,t-2d,t-1d,t\big]$.
\end{itemize}

We use the time window to forecast the flares existence some periods ahead of the $t$ instant (forecasting times or horizons). Our target feature is defined as the occurrence of at least one M- or X-class flare event within the next 24, 24-48, 48-72, and 72-96 hours ahead of the $t$ instant\footnote{Data prepared in this section is available from https://doi.org/10.5281/zenodo.2597637}. Besides, as the NOAA/SWPC DSD dataset holds aggregated flare data (i.e., the amount of flares in each class summed across all numbered sunspot regions), our models will provide binary full-disk flare forecasts regardless of which active region produced the event.

This approach for previewing flare events is similar to what The Met Office Space Weather Operations Centre (MOSWOC) at the United Kingdom employs \citep{Murray:2017}. Using a hybrid-forecasting technique, they make full-disk forecasts based on the Poisson probabilities of the \cite{McIntosh:1990}'s classes and predictions are adjusted by human experts. However, they predict specific classes of flares in the next 24, 24-48, 48-72, and 72-96 hours instead of flares above a magnitude threshold as in this paper. 

The horizons we proposed also matched the set up used by NOAA/SWPC, except for 72-96 hours \citep{Crown:2012}. Besides, as argued by \cite{Jonas:2018}, predicting if there will be a flare event in a certain period ahead of the $t$ instant is a much more realistic problem than predicting flares exactly after $t$.

\section{Proposed Framework}\label{sec:proposedFramework}
After dealing with those data preparation issues, it is worth talking about how we designed our framework to support the flare forecasting systems design.
 
\begin{figure}
\centering
\includegraphics[width=.475\textwidth]{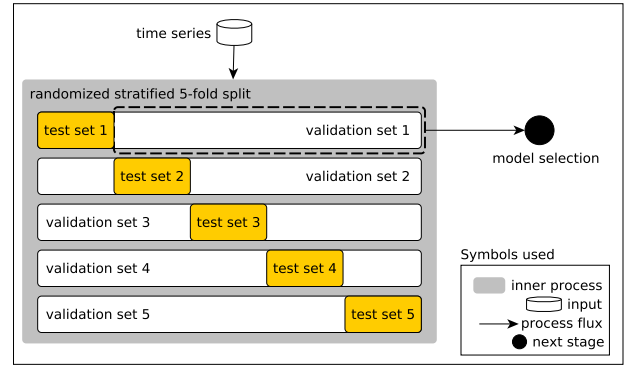}
\caption{Test sets reservation schema.}
\label{fig:schema-test-sets-reservation}
\end{figure}

The first framework process comprehends a data split into test and validation sets (data that will be further investigated). At the beginning, we perform a randomized, stratified 5-fold split according to Figure~\ref{fig:schema-test-sets-reservation}, which guarantees an equally distributed positive and negative classes ratio over all the test and validation sets. We also put the test sets aside and never use them again until the end, when the output model has its prediction error assessed, which is its generalization error over unseen data. Splitting data into 5 folds is common in literature since it reduces the predictors variance and consequently leverage their performance \citep{Hastie:2009}.

According to Figure~\ref{fig:schema-test-sets-reservation}, we perform the next framework stage (model selection) over each validation set individually, one at a time. After model selection, several other processes are carried out in each validation set: feature selection, hyper-parameters optimization, and data resampling. We use the output of each process to decide which aspect we will have in the final classifier (features, algorithm, fine-tuning, and other aspects). 

\subsection{Model Selection}
After dealing with the initial splitting of subsets, the focus is to pick models that best fit data and can minimize the validation error without major adjustments, namely the model selection process. 

\begin{figure*}
\centering
\includegraphics[width=0.8\textwidth]{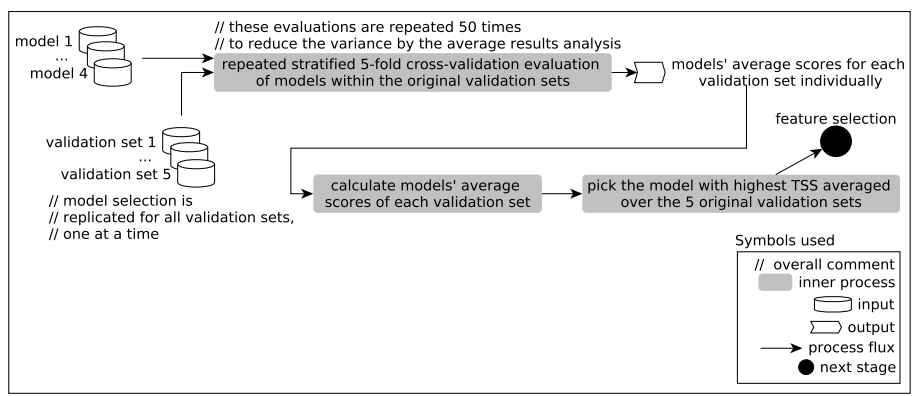}
\caption{Model selection schema.}
\label{fig:model-selection-schema}
\end{figure*}

We now evaluate some distinct machine learning algorithms on validation sets, as shown in Figure~\ref{fig:model-selection-schema}. We assess their performance through a repeated randomized, stratified 5-fold cross-validation strategy over each validation set. The repetition of the k-fold cross-validation is used to produce a more reliable estimate of performance, since it reduces the data variance (the mean of all iterations gives the overall performance)~\citep{Hastie:2009}. 

The criterion is to pick the algorithm that maximizes the true skill statistics (TSS) quality score \citep{Jolliffe:2003} among all validation sets (Figure~\ref{fig:model-selection-schema}). This score ranks the model  performance over a scale ranging from $-1$ (all predictions incorrect) to $1$ (all predictions correct). As shown in Equation~\ref{eq:tss}, the TSS combines the positive and negative outcome classes individual success rates, which is considered one of its strengths. 

\begin{equation}\label{eq:tss}
\begin{gathered}
TSS = TPR + TNR - 1 
\end{gathered}
\end{equation}
\noindent where TPR \citep{Han:2006,Jolliffe:2003} is the true positive rate -- also known as recall -- and measures the number of positive samples correctly predicted. In turn, the TNR is the true negative rate and accounts for the number of negative samples correctly predicted \citep{Han:2006,Jolliffe:2003}. Equation~\ref{eq:sensitividade} shows how to calculate the TPR.

\begin{equation}\label{eq:sensitividade}
\begin{gathered}
TPR = \frac{TP}{TP+FN} 
\end{gathered}
\end{equation}
\noindent where TP refers to the true positives (positive samples predicted as positive) and FN comprehends the false negatives (positive samples that were incorrectly classified) \citep{Han:2006}. We measure the TPR on a scale ranging from 0 to 1 where higher values are better. Similarly to the TPR, the TNR is also scored in the scale \big[0,1\big], as Equation~\ref{eq:especificidade} shows.

\begin{equation}\label{eq:especificidade}
\begin{gathered}
TNR = \frac{TN}{TN+FP} 
\end{gathered}
\end{equation}
\newline
\noindent where TN refers to the true negatives (negative samples predicted as negative) and FP corresponds to the false positives (negative samples that were incorrectly classified) \citep{Han:2006}.

\subsubsection{Decision Trees Ensembles}

The models trained during model selection are ensembles, i.\,e. algorithms that combine the output of several individual models to optimize the predictive performance. Since some classifiers are better than others in specific scenarios, it is rather important to have some cooperation between them to minimize the noise of weak models \citep{Witten:2011}. Ensembles reduce the data variance when combined with the bagging strategy, which is complementary to the repeated k-fold cross-validation strategy in reducing variance \citep{James:2013}. 

We evaluated four types of decision trees ensembles, all of them with bagging as the sampling strategy: gradient tree boosting \citep{Hastie:2009}, Ada\-Boost \citep{Zaki:2013}, random forest \citep{James:2013} and bagging classifier \citep{Han:2006}. At the beginning, the base learners are 60 classification and regression decision trees (CART) for all ensembles \citep{Breiman:1984}.

The bagging classifier is a meta-algorithm that creates several individuals of the same type for an ensemble by training each of them on a different sampling of data (bagging strategy). By sampling with replacement, this method allows to reduce variance and consequently increases performance. The individual classifiers results combination can be done according to a majority voting strategy (hard voting) or by averaging the individual probabilities outputs (soft voting).

The random forest classifier is also a meta-algorithm that fits several individual classifiers (decision trees) on distinct data samples drew through the bagging strategy. It uses soft voting to combine the trees probabilities outputs to improve the predictive performance while controlling over-fitting \citep{James:2013}. Other authors that effectively used random forest in flares forecasting are \cite{Liu:2017c}.

Unlike the two aforementioned meta-algorithms that independently fit several models and aggregate the results at the end without preference for any individual classifier, boosting is a strategy in which every individual model drives the samples on which the next models will focus. The AdaBoost is the algorithm that introduced the boosting strategy \citep{Zaki:2013}. 

AdaBoost comprehends a meta-estimator that begins by fitting a single learner on the original dataset, thus fitting additional copies of this learner focusing on the incorrectly classified samples. Hence, AdaBoost focuses on the most challenging samples \citep{Zaki:2013}. Other authors that effectively used AdaBoost to forecast flares are \cite{Lan:2012}.

In turn, the gradient tree boosting is another algorithm that relies on the boosting strategy. This model also trains several individuals gradually and sequentially. The main difference between AdaBoost and a gradient tree boosting schema is while the former identifies the weakness of base classifiers by adjusting weights in samples hard to predict, the latter optimizes its loss function \citep{Hastie:2009} -- in this research, a deviance loss function. To the best of our knowledge, this is the first time gradient tree boosting models are employed in flares forecasting.

The rationale for using decision trees instead of other robust models like neural nets \citep{Witten:2011} or support-vector machines (SVMs) \citep{Zaki:2013} is mainly because they have a natural skill for handling mixed data and have robustness for outliers \citep{Hastie:2009}. In general, while the predictive performance of trees may be slightly worse than neural nets and SVMs, we manage to handle this disadvantage with the use of ensembles.

\subsection{Feature Selection}
\label{sec:feature-selection}
After dealing with the selection of a model that best fits data without significant refinements, the framework picks the features that best contribute to the previously chosen model. This section will further describe which aspects we have to cope with during the feature selection process.

\begin{figure*}
\centering
\includegraphics[width=1.0\textwidth]{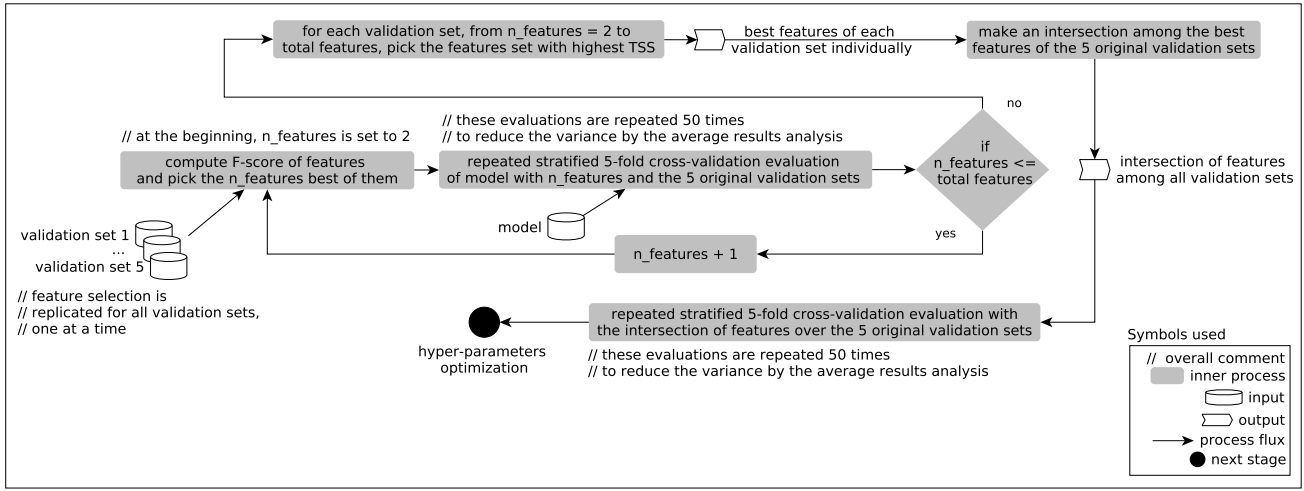}
\caption{Feature selection schema.}
\label{fig:feature-selection-schema}
\end{figure*}

The chosen ensemble is re-evaluated on the validation sets using the same repeated randomized, stratified 5-fold cross-validation strategy. However, some analyses on selecting and discarding features are carried out before those re-evaluations, as shown in Figure~\ref{fig:feature-selection-schema}.

Mostly because solar flares forecasting systems must deal with right input data, designing efficient and accurate models is not usually an easy task. Moreover, when working with such systems, one must pay careful attention to the input data since inputting noisy or useless features into those systems can lead to poor results (poor system performance) \citep{Han:2006}. Besides, getting rid of weak features is also a natural way of reducing the data input dimensionality that decreases the learner performance. Thus, feature selection analyses are useful tasks used to select only the best attributes

We chose two strategies to measure features usefulness and support discarding some of them: a filter method combined with a wrapper schema \citep{Guyon:2003}. While the former uses a proxy measure to score features subsets and creates ranks of features importance (for instance, the Pearson's correlation coefficient \citep{Han:2006}), the latter uses a predictive model to score those sets (for instance, several features subsets are used to train a baseline model and have their performance evaluated). In this paper, we used a univariate feature selection schema provided with the F-score calculation \citep{Bobra:2015} as the filter method and the chosen ensemble as the wrapper.

\subsubsection{Univariate Feature Selection}

Univariate feature selection schemas do not account for the correlation between the input space elements during their analyses. Instead, they assume that features are independent at the same time they test their complimentary nature. This nature means that one feature may be a poor predictor alone; however, it can be strong if combined with others. Thus, complimentary features do not necessarily comprehend a correlated relationship~\citep{Guyon:2003}. 

Besides, as shown by \cite{Bobra:2015}, the flare predictors literature had already proved the benefits of selecting features through univariate schemas provided with the F-score calculation. Equation~\ref{eq:f-score} shows how to calculate the F-score \citep{Chang:2008}.

\begin{equation}\label{eq:f-score}
\begin{gathered}
F(i)=\frac{(\bar{x}_{i}^{+}-\bar{x}_{i})^{2}+(\bar{x}_{i}^{-}-\bar{x}_{i})^{2}}{\frac{1}{n^{+}-1}\sum_{k=1}^{n^{+}}(x_{k,i}^{+}-\bar{x}_{i})^{2}+\frac{1}{n^{-}-1}\sum_{k=1}^{n^{-}}(x_{k,i}^{-}-\bar{x}_{i})^{2}}
\end{gathered}
\end{equation}

\noindent where the numerator refers to the inter-class variance; the denominator corresponds to the variance sum within each class; $\bar{x}_{i}^{-}$ and $\bar{x}_{i}^{+}$ are the negative and positive samples mean values, respectively; $\bar{x}_{i}$ corresponds to the samples mean values; and $n^{-}$ and $n^{+}$ are the number of negative and positive samples, respectively. 

For each validation set, we calculate the F-score of all 55 features and organize the output rank in descendant order, as shown in Figure~\ref{fig:feature-selection-schema}. Then, we pick the two best elements available and evaluate the chosen ensemble with them through repeated randomized, stratified 5-fold cross-validation. We repeat this process from the two best features until their total amount, always increasing the number of the picked best features by 1 and recording the TSS of each iteration.

This approach allows us to pick the five best features sets over all the validation sets at the end, i.e., the ones which led to the highest TSS scores. Then, we intersect between the elements in those five best features sets and keep the ones that are common to all of them. Finally, we re-evaluate the chosen ensemble over all the validation sets using this intersection of features through repeated randomized, stratified 5-fold cross-validation.

\subsection{Hyper-parameters Optimization}
\label{sec:hyper-parameters Optimization}

\begin{figure*}
\centering
\includegraphics[width=1\textwidth]{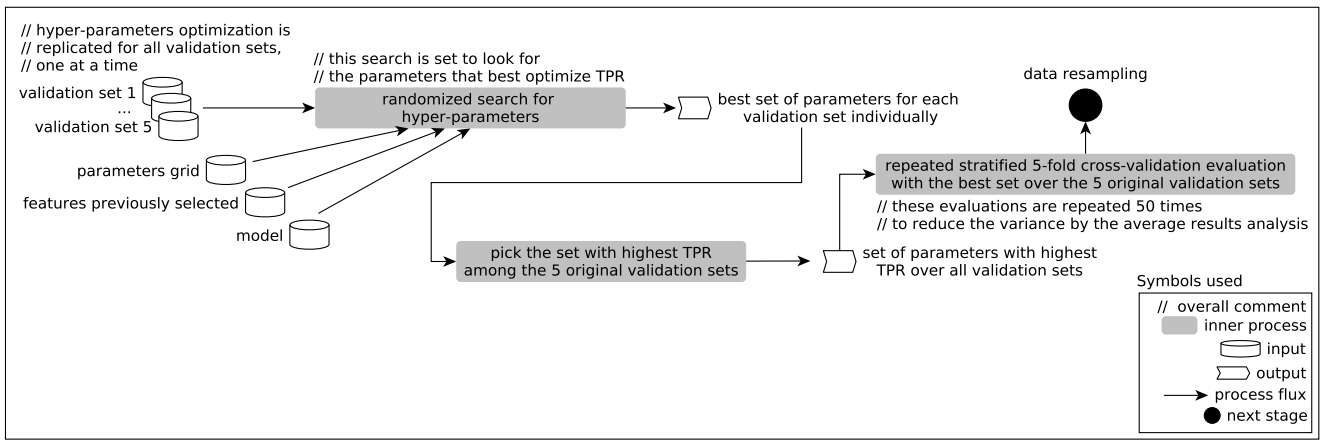}
\caption{Hyper-parameters optimization schema.}
\label{fig:hyper-parameters-optimization-schema}
\end{figure*}

This section describes which aspects we have to cope with during the hyper-parameters optimization process. Given that we have an affordable features subset, it is now time to adjust how the chosen ensemble behaves to better leverage its generalization skills.

A typical learning algorithm aims to train a model $\mathcal{M}$ that minimizes a loss function $\mathcal{L}(X^{(val)};\mathcal{M})$ on $X^{(val)}$ validation data samples. Typical loss functions include the mean squared error for regression problems and the error rate for the classification ones. Model $\mathcal{M}$ is fit with an algorithm $\mathcal{A}$ and some $X^{(tr)}$ training data. Usually, this algorithm $\mathcal{A}$ is provided with a $\lambda$ hyper-parameters set, $\mathcal{M}=\mathcal{A}(X^{(tr)};\lambda)$, commonly optimized (adjusted) to reduce the loss function \citep{Claesen:2015}. 

Besides minimizing its loss function, the model $\mathcal{M}$ must have an appropriate complexity level. Complex models poorly generalize unseen data (over-fitting). On the other hand, excessively simple models may not appropriately learn the data patterns (under-fitting). 

The model complexity also affects the bias-variance trade-off,~i.e., complex models usually have high variance, while the simple ones are somehow biased. Therefore, learning algorithms also provide their hyper-parameters in an attempt to balance the bias-variance trade-off \citep{Claesen:2015}. Nonetheless, hyper-parameters are usually involved in a search process that seeks a set of them ($\lambda^{\star}$) that yield an optimal model, as Equation~\ref{eq:hyper-parameter-tuning} defines \citep{Claesen:2015}.


\begin{equation}\label{eq:hyper-parameter-tuning}
\begin{split}
\lambda^{\star} & = \underset{\lambda}{\mathrm{arg \ min}} \ \mathcal{L}(X^{(val)};\mathcal{A}(X^{(tr)};\lambda)) \\
    & = \underset{\lambda}{\mathrm{arg \ min}} \ \mathcal{F}(\lambda; \mathcal{A},X^{(tr)},X^{(val)}, \mathcal{L})
\end{split}
\end{equation}

\noindent where the $\mathcal{F}$ function uses some hyper-parameters and outputs the loss value provided that the $X^{(tr)}$ training data, the $X^{(val)}$ validation data and the $\mathcal{L}$ loss function are given.

\subsubsection{Randomized Hyper-parameters Search}

Common techniques for carrying out hyper-parameters search include the grid search and the random search. While the former takes a grid of parameters along with their corresponding data ranges and searches all the combinations, the latter also takes this grid; however, the parameters values are randomly sampled over the available ranges. 

Although simple to code and effective, grid search is computationally expensive when there are high dimensional input spaces -- as this paper. Random search, in turn, can be as effective as the grid-based technique. Random-based searches are equally efficient since not all parameters are important to tune and not all input dimensions are useful to search~\citep{Bergstra:2012}. Hence, we proposed the use of a random search process provided with the following hyper-parameters grid: 

\begin{itemize}
    \item the ensembles inner trees amount;
    \item the ensembles inner trees max depth;
    \item the samples minimum amount required to split inner trees nodes;
    \item the samples minimum amount needed at a leaf node;
    \item the features amount used when looking for the inner trees best splits;
    \item the samples amount used to fit the inner trees (bagging). 
\end{itemize}

The aforementioned parameters are common to all the considered ensembles. Also, we carry out the hyper-parameters search to minimize the error rate when forecasting the positive samples. This approach consequently increases the recalls of models.

As Figure~\ref{fig:hyper-parameters-optimization-schema} shows, we perform a randomized search over the validation sets to look for the hyper-parameters set up that most increases the recall in each of them. Here, we use recall instead of TSS, since increasing the former is a way to naturally increase the latter.

Then, we pick the parameters optimal set (the one that most increased recall over all the validation sets) and re-evaluate the chosen ensemble provided with this set over each validation set using repeated randomized, stratified 5-fold cross-validation.

\subsection{Data Resampling}
\label{sec:data-resampling}
Besides choosing the appropriate features and adjusting the chosen ensemble complexity, one should also consider the scenario of fitting learning models in imbalanced datasets. This section will describe the rationale behind dealing with this issue and also how we manage to minimize its effects.

\begin{figure*}
\centering
\includegraphics[width=0.60\textwidth]{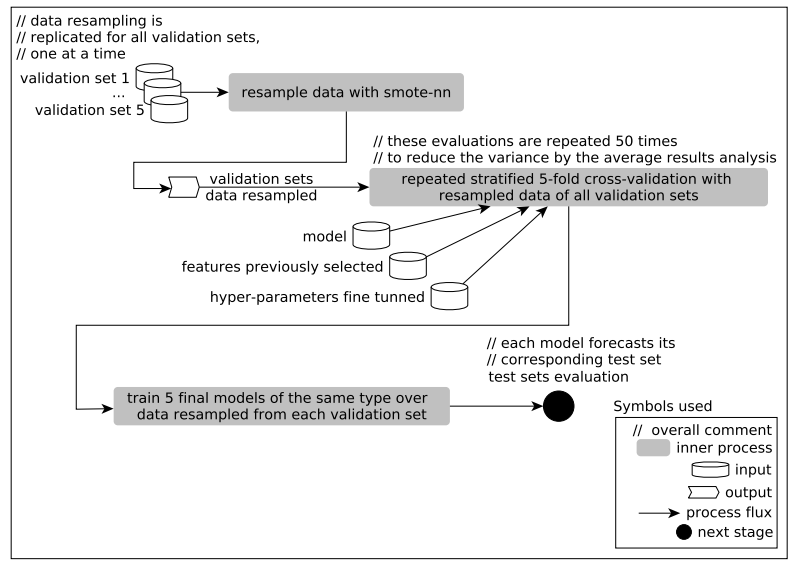}
\caption{Data resampling schema.}
\label{fig:data-resampling-schema}
\end{figure*}

Usually, datasets used in real-world forecasting problems have imbalanced class ratios. Negative samples are much more common than positive ones. In our dataset, before splitting the test sets, there was approximately 17.47\% of data representing the positive class (i.e., 1279 examples from January 01, 1997 to January 15, 2017 had at least one M- or X-class flare occurring) while the rest accounted for the negative class (6041 examples, 82.53\%).

Fitting models on imbalanced data have a high cost and create over-fitted classifiers concerning the majority class, i.e., classifiers that cannot generalize well minority samples, thus only being able to forecast majority samples \citep{Chawla:2002}. A straightforward way to deal with imbalanced data is to change how classifiers report their performance. Instead of only using the biased accuracy score to measure the classifier effectiveness, one can use individual metrics to verify performance when forecasting the individual classes (for instance, the positive -- TPR -- and negative -- TNR -- class hit rate) \citep{Batista:2004}. 

The verification of the individual performance may not be enough when we also want to increase predictive performance. The literature also addresses the class imbalance issue in other ways, such as by assigning penalty costs to training examples \citep{Bobra:2015} and by resampling the dataset \citep{Chawla:2002} -- as we did in this paper.

\subsubsection{Combined Method for Data Resampling}

We chose the SMOTE-ENN technique \citep{Batista:2004} to deal with the imbalanced data ratios in our flares catalog. This approach combines the Synthetic Minority Over-sampling Technique (SMOTE) \citep{Chawla:2002} with the Wilson’s Edited Nearest Neighbor (ENN) rule \citep{Wilson:1972}.

Instead of randomly duplicating with replacement the minority class samples as ordinary random over-sampling methods, SMOTE creates synthetic  minority class samples by interpolating between positive class samples that lie together \citep{Chawla:2002}. This interpolation helps spread the positive class decision boundary over the majority class space, so the bias produced by a simple random over-sampling duplication with replacement is reduced~\citep{Batista:2004}.

Regarding SMOTE-ENN, the dataset is first over-sampled with SMOTE, and then the ENN rule is carried out over it \citep{Batista:2004}. Conversely to SMOTE, ENN is an under-sampling technique. Instead of randomly deleting majority class samples that can lead to important data loss, ENN removes samples from the training set that are misclassified by its 3-nearest-neighbors algorithm. 

As Figure~\ref{fig:data-resampling-schema} shows, we resample each validation set with SMOTE-ENN, and the output sets are used to fit the optimized set up from Section~\ref{sec:hyper-parameters Optimization}. Thus, we create five models of the same type (i.e., the same algorithm, features, and hyper-parameters), fitted over each individual validation set. Those models are used to forecast their corresponding test sets reserved at the beginning. 

At this point, it is worth noting that, in this work, we have empirically defined the pipeline order of inner processes. We are aware that the final solution may not be the optimal classifier (i.e., changing the order we carry out feature selection, hyper-parameters optimization, and data resampling might somehow alter the output model), but at least the proposed pipeline leads to a high-quality one. 

Obtaining the optimal classifier could be possible, provided that we try all possible pipeline arrangements. However, the nature of this search would be exhaustive and probably unfeasible in practice.

In fact, the rationale for carrying out data resampling, in the end, refers to SMOTE-ENN, which creates a lot of synthetic samples -- despite having an under-sampling step -- and this significantly increases the dataset size. If we have considered resampling in the beginning, other processes could become significantly slower, especially because of the repeated randomized, stratified 5-fold cross-validations. 

However, as we treat our inner processes as blocks with well-defined inputs and outputs, changing the proposed order of processes to other desired arrangement could be done with little effort. To justify our rationale, we can cite the research by \cite{Zhang:2017}, which observed some effects of inverting machine-learning design processes.

In this sense, \cite{Zhang:2017} investigated what happens when a resampling approach -- in their case, under- and over-sampling -- is used before feature selection and vice versa. They designed experiments with nine feature selection methods, six resampling approaches, three well-known classifiers, and 35 datasets. 

Accordingly, \cite{Zhang:2017} compared the performance of their models in terms of accuracy, balanced accuracy, and f1-scores. Overall, they concluded that there is not any winner between both process arrangements. In essence, researchers can test both to derive the best classifiers (sometimes the earlier use of data resampling outperforms the later feature selection and vice versa). This conclusion corroborates our argument for suggesting other pipeline arrangements when needed.

\subsection{Test Sets Evaluation}
\label{sec:test-set-evaluation}
Besides forecasting with data resampling output models on the test sets, we also fit five baseline models over the validation segments before their resampling and forecast their corresponding test sets. The baseline models have the same set up as the chosen ensemble during the model selection process. We fit the baseline models to verify the framework decision-making process effectiveness, i.e., how the predictive performance changed from the baseline to the framework proposed intervention.

\section{Results}\label{sec:Results}
This section will present the results of each framework stage, emphasizing how performance changed. In addition to TPR, TNR, and TSS, we will use other metrics to report the performance, such as the area under the receiver operating characteristic (ROC) curve (AUC) \citep{Zaki:2013} and accuracy (ACC) \citep{Han:2006}.

The AUC is a score that measures the two-dimensional area underneath the ROC curve \citep{Zaki:2013}. The ROC analysis plots the TPR score (y-axis) against the false positive rate (FPR) (x-axis) \citep{Zaki:2013} for increasing decision thresholds. The calculated area is always positive and ideally should be greater than~0.5. 

Also called the probability of a false detection (POFD), the FPR calculates the probability of detecting false alarms among the negative predictions. Equation~\ref{eq:fpr} shows how to calculate the FPR.

\begin{equation}\label{eq:fpr}
FPR = \frac{FP}{TN + FP} 
\end{equation}
\noindent where FP is the number of false positives and TN is the number of true negatives. 

By combining TPR and FPR, the AUC represents the classifiers skill to discriminate between positive and negative events. Thus, it can be used to indirectly analyze the number of false alarms predicted, i.e., the higher the probability of those elements occur, thus leading to higher FPRs, the more the AUC score value decreases. 

Therefore, best classifiers score the AUC next to the left-hand corner (FPR = 0 and TPR = 1). On the other hand, worst classifiers score next to the bottom right-hand corner (FPR = 1 and TPR = 0).

The accuracy, in turn, is the well-known metric that accounts for the correct forecasts (positive or negative) divided by the total amount of them. Unfortunately, ACC only reports reliable results when there are balanced class ratios; however, we chose to keep this metric for completeness since most researchers use it.

\subsection{Model Selection Results}
Table~\ref{tab:model-selection-results} shows the results from model selection. Results refer to the mean scores values among all validation sets and cover all forecasting periods we designed. 

Overall, we can see reasonable ACC values along with high TNR values and low TPR values for all models and forecasting periods. This is common when there are extremely imbalanced data samples. It means that classifiers are not able to correctly forecast the positive class while they perform well with the negative one, which increases ACC by definition. 

\begin{table}
\centering
\caption{Model selection results.}
\label{tab:model-selection-results}
\scalebox{0.8}{
\begin{tabular}{ccccccc}
\hline
\multicolumn{1}{c}{\textit{Forecasting Time}} & \textit{Model}       & \textit{ACC} & \textit{TPR} & \textit{TNR} & \textit{AUC} & \textit{TSS} \\ \hline
\multirow{4}{*}{next 24 hours}     & Bagging              & 0.85         & 0.25         & 0.97         & 0.85         & 0.22         \\
                                              & RandomForest         & 0.84         & 0.25         & 0.97         & 0.85         & 0.22         \\
                                              & AdaBoost             & 0.84         & 0.34         & 0.95         & 0.84         & 0.29         \\
                                              & GradientTreeBoosting & 0.85         & 0.35         & 0.95         & 0.85         & 0.30         \\ \hline
\multirow{4}{*}{24-48 hours}  & Bagging              & 0.84         & 0.20         & 0.98         & 0.83         & 0.18         \\
                                              & RandomForest         & 0.84         & 0.21         & 0.97         & 0.83         & 0.18         \\
                                              & AdaBoost             & 0.84         & 0.29         & 0.95         & 0.82         & 0.24         \\
                                              & GradientTreeBoosting & 0.84         & 0.30         & 0.96         & 0.83         & 0.26         \\ \hline
\multirow{4}{*}{48-72 hours}  & Bagging              & 0.84         & 0.15         & 0.98         & 0.81         & 0.13         \\
                                              & RandomForest         & 0.84         & 0.16         & 0.98         & 0.81         & 0.14         \\
                                              & AdaBoost             & 0.83         & 0.23         & 0.96         & 0.80         & 0.19         \\
                                              & GradientTreeBoosting & 0.83         & 0.24         & 0.96         & 0.81         & 0.20         \\ \hline
\multirow{4}{*}{72-96 hours}  & Bagging              & 0.83    -     & 0.13         & 0.98         & 0.80         & 0.11         \\
                                              & RandomForest         & 0.83         & 0.14         & 0.98         & 0.80         & 0.12         \\
                                              & AdaBoost             & 0.83         & 0.21         & 0.96         & 0.79         & 0.17         \\
                                              & GradientTreeBoosting & 0.83         & 0.22         & 0.96         & 0.80         & 0.18         \\ \hline
\end{tabular}}
\end{table}

We can also observe the false positives occurrence in the AUC score since all models achieved values close to 0.8. Provided that the ROC curve will generate a perfect AUC = 1 when FPR values are next to zero, we can see that false alarms did occur in some way since the AUCs decreased on average by about 0.22. Besides, the gradient tree boosting models scored the best TSS values for all forecasting horizons, and thus were kept through next stages.

Finally, Table~\ref{tab:model-selection-results} also shows decreasing score values among all classifiers as we increase the forecasting time. For instance, the gradient tree boosting model TSS score decreased by 0.04 when the forecasting horizon changed from next 24 to 24-48 hours and by 0.06 for 48-72 hours. Nevertheless, this is an expected behavior since we try to preview flares more days ahead and the difficulty substantially increases, thus damaging the models performance.

\subsection{Feature Selection Results}

Table~\ref{tab:feature-selection-results} shows the feature selection results. Results refer to the features intersection set (Section~\ref{sec:feature-selection}) re-evaluated over all the validation sets.

Compared with the model selection results shown in Table~\ref{tab:model-selection-results}, we can see subtle increases of TSS (next 24 and 72-96 hours) and TPR (24-48, 48-72, and 72-96 hours). Overall, the feature selection increased these metrics by 0.01. 

\begin{table}
\centering
\caption{Feature selection results.}
\label{tab:feature-selection-results}
\resizebox{0.45\textwidth}{!}{
\begin{tabular}{ccccccc}
\hline
\textit{Forecasting Time}                       & \textit{Model}                           & \textit{ACC}             & \textit{TPR}             & \textit{TNR}             & \textit{AUC}             & \textit{TSS}             \\ \hline
next 24 hours                        & GradientTreeBoosting                     & 0.85                     & 0.35                     & 0.95                     & 0.85                     & 0.31                     \\
24-48 hours                     & GradientTreeBoosting                     & 0.84                     & 0.31                     & 0.96                     & 0.83                     & 0.26                     \\
48-72 hours                     & GradientTreeBoosting                     & 0.83                     & 0.25                     & 0.96                     & 0.81                     & 0.20                     \\
72-96 hours & GradientTreeBoosting & 0.83 & 0.23 & 0.96 & 0.80 & 0.19 \\ \hline
\end{tabular}}
\end{table}

Although increases may look small at first, it is worth emphasizing that besides increasing the predictive performance, feature selection also reduces the input dimensionality since it discards features that do not contribute in a significant manner. Therefore, feature selection is rather important in the framework pipeline since it defines the sliding time window length of features while adjusting models bias by discarding irrelevant elements.

\subsection{Hyper-parameters Optimization Results}
Table~\ref{tab:hyper-parameter-optimization-results} shows the hyper-parameters optimization results. Results refer to the parameters set that best increased the TPR (Section~\ref{sec:hyper-parameters Optimization}) re-evaluated over all the validation sets.

Compared with the feature selection results shown in Table~\ref{tab:feature-selection-results}, we can see increases in the TPR absolute values over all the forecasting horizons. For instance, the highest TPR increases were 0.08 for 72-96, and 0.07 for 48-72 hours.

\begin{table}
\centering
\caption{Hyper-parameters optimization results.}
\label{tab:hyper-parameter-optimization-results}
\resizebox{0.45\textwidth}{!}{
\begin{tabular}{ccccccc}
\hline
\textit{Forecasting Time}   & \textit{Model}       & \textit{ACC} & \textit{TPR} & \textit{TNR} & \textit{AUC} & \textit{TSS} \\ \hline
next 24 hours    & GradientTreeBoosting & 0.82         & 0.41         & 0.91         & 0.80         & 0.32         \\
24-48 hours & GradientTreeBoosting & 0.81         & 0.36         & 0.90         & 0.77         & 0.26         \\
48-72 hours & GradientTreeBoosting & 0.80         & 0.32         & 0.91         & 0.75         & 0.23         \\
72-96 hours & GradientTreeBoosting & 0.80         & 0.31         & 0.91         & 0.76         & 0.22         \\ \hline
\end{tabular}}
\end{table}

Besides increasing the TPR values, the hyper-parameters optimization also increased the TSS scores (except for 24-48 hours). The TSS values increased despite the TNR scores decreases over all the forecasting horizons. We also verified some decreases in the AUC scores, which means that the false positives amounts were slightly higher despite the higher TPR values. 

Despite the benefits, there is still room for further improvements. For instance, the model still has high TNR scores compared with the TPR ones, which is explained by the imbalanced data nature that we did not previously cope with. We will make the improvements regarding the resampling of class ratios by the next and last framework stage.

\subsection{Data Resampling Results}
Table~\ref{tab:data-resampling-results} shows the data resampling results. These results refer to the chosen ensemble evaluated over each resampled validation set along with the fine-tuned parameters and features previously selected.

\begin{table}
\centering
\caption{Data resampling results.}
\label{tab:data-resampling-results}
\resizebox{0.45\textwidth}{!}{
\begin{tabular}{ccccccc}
\hline
\textit{Forecasting Time}   & \textit{Model}       & \textit{ACC} & \textit{TPR} & \textit{TNR} & \textit{AUC} & \textit{TSS} \\ \hline
next 24 hours    & GradientTreeBoosting & 0.91         & 0.93         & 0.88         & 0.93         & 0.81         \\
24-48 hours & GradientTreeBoosting & 0.92         & 0.94         & 0.88         & 0.95         & 0.82         \\
48-72 hours & GradientTreeBoosting & 0.94         & 0.97         & 0.89         & 0.98         & 0.86         \\
72-96 hours & GradientTreeBoosting & 0.94         & 0.98         & 0.89         & 0.98         & 0.86         \\ \hline
\end{tabular}}
\end{table}

The models now do not suffer from over-fitting regarding the original majority class. The ACC is high for all forecasting periods, and so are the TPR and TNR. Since both TPR and TNR are high, the models also increased their TSS scores.

Besides, the AUC values show the false positive amount decrease since these scores achieved values close to 1 for next 24, 24-48, 48-72, and 72-96 hours. We can also see lower TNR values compared with TPR (before data resampling, the TNR values were higher than the TPR ones). This happens since SMOTE causes a natural trend inversion in learning models, and this consequently reverses the bias in favor of the original minority class~\citep{Chawla:2002}. 

Despite the trend inversion, the TPR and TNR values are high and show that models can now predict flare or non-flare events with high confidence over all validation scenarios. Therefore, we can use the refined models fit with data resampling to forecast unseen samples of test data.

\subsection{Test Set Results}

In addition to forecasting with data resampling output models on test sets, we also verified the predictive performance of five baseline models as described in Section~\ref{sec:test-set-evaluation}.  Table~\ref{tab:pre-framework-test-set-results} shows the results of the baseline predictions while Table~\ref{tab:post-framework-test-set-results} shows the results from the framework output models.

\begin{table}
\centering
\caption{Baseline models evaluation over test sets results.}
\label{tab:pre-framework-test-set-results}
\resizebox{0.45\textwidth}{!}{
\begin{tabular}{ccccccc}
\hline
\textit{Forecasting Time}   & \textit{Model}       & \textit{ACC} & \textit{TPR} & \textit{TNR} & \textit{AUC} & \textit{TSS} \\ \hline
next 24 hours    & GradientTreeBoosting & 0.85         & 0.35         & 0.95         & 0.85         & 0.30         \\
24-48 hours & GradientTreeBoosting & 0.84         & 0.30         & 0.96         & 0.83         & 0.26         \\
48-72 hours & GradientTreeBoosting & 0.83         & 0.23         & 0.96         & 0.81         & 0.19         \\
72-96 hours & GradientTreeBoosting & 0.83         & 0.21         & 0.97         & 0.80         & 0.18         \\ \hline
\end{tabular}}
\end{table}

\begin{table}
\centering
\caption{Framework output models evaluation over test sets results.}
\label{tab:post-framework-test-set-results}
\resizebox{0.45\textwidth}{!}{
\begin{tabular}{ccccccc}
\hline
\textit{Forecasting Time}   & \textit{Model}       & \textit{ACC} & \textit{TPR} & \textit{TNR} & \textit{AUC} & \textit{TSS} \\ \hline
next 24 hours    & GradientTreeBoosting & 0.71         & 0.75         & 0.70         & 0.79         & 0.46         \\
24-48 hours & GradientTreeBoosting & 0.65         & 0.70         & 0.64         & 0.71         & 0.34         \\
48-72 hours & GradientTreeBoosting & 0.70         & 0.75         & 0.69         & 0.79         & 0.44         \\
72-96 hours & GradientTreeBoosting & 0.69         & 0.73         & 0.68         & 0.78         & 0.42         \\ \hline
\end{tabular}}
\end{table}

We observe increases in the TPR scores, which raised from 0.35, 0.30, 0.23, and 0.21 to 0.75, 0.70, 0.79, and 0.78, respectively, for the next 24, 24-48, 48-72, and 72-96 hours. Followed by the TPR increased values, the TSS ones also raised considerably.

The TNR, in turn, decreased over all the scenarios, mostly because the models trend inversion caused by SMOTE previously described. This behaviour also indirectly explains why the TSS scores did not have significant increases as the TPR ones. Since the TSS can be understood as a direct relation between TPR and TNR, both of them must be high at the same time to push the TSS to higher levels.

The models kept the AUC scores values close to the baseline, which means that the false positive numbers were kept at similar levels, except for 24-48 hours that had the AUC decreased by 0.12. It is noteworthy that the framework did produce significant positive effects in the models skill to forecast the positive class while it did not notably change their precision, which would result in more false alarms. 

Besides, we managed to increase TPRs without touching on algorithms' decision thresholds. Adjusting those thresholds would naturally increase TPR but also produce a high number of false positives (the false alarm ratio (FAR), as defined by \cite{Jolliffe:2003}), thus decreasing AUC, as several authors are used to do to leverage their recall. Thus, we kept our decision thresholds constant during the whole design process, noticeably at the default level (0.5).

Finally, by comparing the relationship between ACC, TPR, and TNR in Tables \ref{tab:pre-framework-test-set-results} and \ref{tab:post-framework-test-set-results}, we can see that models changed from over-fitted versions that favored the original negative class to versions that could  forecast the positive or negative class without any preference.

\section{Literature Analysis} 
\label{sec:comparisons-with-literature}
In this section, we will position our forecast performance within the literature. To further detail the involved papers, we shall describe how authors designed their systems, emphasizing how they assembled their dataset and features, and estimated their prediction errors under unseen data (Sections \ref{sec:not-operationally-evaluated Systems} and \ref{sec:operationally-evaluated Systems}).

\cite{Barnes:2008} and \cite{Barnes:2016} concluded that, unless the datasets are identical, there is not enough meaning in comparing metrics from different methods directly. Examples of characteristics that prevent direct comparisons include different data segmentation strategies into training and test sets, how authors designed their forecasts (target features), and also the prediction type (i.e., whether it is full-disk or AR-by-AR).

The characteristics mentioned earlier introduce uncertainty when directly comparing TSS and other metrics from distinct models. It does not become clear whether observed differences in results are due to used methods or to differing datasets~\citep{Barnes:2008, Barnes:2016}. 

Furthermore, we shall drive this section to separate systems between two distinct groups: operationally-evaluated (evaluated without any bias) and not-operationally-evaluated (evaluated under some bias). In fact, biased results are not wrong, but they refer to systems evaluated under specific scenarios that can mask their real generalization skills in real operational settings. Henceforth, let us use the following four criteria to distinguish between operationally- and not-operationally-evaluated approaches: 

\begin{enumerate}
    \item \emph{Evaluation without truly unseen data}: as argued by \cite{Ahmed:2013}, several flare forecasting authors do not test their systems with real test data, i.e., unseen data. Working on how some data subsets can be removed from the training process gives a true estimate of systems performance while distinguishing between flare and non-flare unseen samples. For instance, \cite{Ahmed:2013} proposed to split their datasets into years of training and testing data. On the other hand, we proposed the framework with the random-based test data splitting right at the beginning, before any treatment used for designing the output model. Hence, proposals that do not explicitly use unseen data to report their results will be considered not-operationally-evaluated.
    \item \emph{Use of magnetograms with ARs near the solar disk center only}: some authors only include magnetograms with ARs near the solar disk center when picking examples for their datasets, i.e., within a defined radius. However, this approach raises uncertainty about the reported scores and weakens the interpretation of their results for operational purposes, since in real operational environments their systems must behave with ARs at any location in the disk, including far from the center (at the limb) \citep{Nishizuka:2017}. Therefore, those proposals shall also be considered not-operationally-evaluated.
    \item \emph{Use of magnetograms with ARs linked to $\geq C1.0$ flares only}: some authors only include magnetograms with ARs producing $\geq C1.0$ flares and distinguish samples linked to $\geq M1.0$ events as the positive class. However, this criterion also raises uncertainty about the reported scores and weakens the interpretation of their results for operational purposes, since in real operational environments their systems must behave with ARs that are not linked to any sort of flare \citep{Bloomfield:2012}. Hence, those proposals will be considered not-operationally-evaluated.
    \item \emph{Lack of enough data during models designing:} some authors only include few data samples while fitting their models, which would affect their systems generalization skill in operational settings, i.e., under-fitting. Models designed with a reduced number of samples are not as effective as those designed over more representative datasets. Adding more samples to datasets leads to representative training sets, thus allowing finer discrimination between feature values \citep{Pyle:1999}. Those proposals shall also be considered not-operationally-evaluated.
\end{enumerate}

In this sense, for results agreement between distinct papers, operationally-evaluated systems will take our test set results as a reference, which are free from any bias, whereas not-operationally-evaluated ones (biased) will use our data resampling results for reference. 

Noteworthily, we will highlight our literature research over the last ten years because we want to give an in-depth look at the state-of-the-art. Nevertheless, we may also include some seminal papers. 

Last, but not least, for scope definition purposes, our literature research shall include:

\begin{itemize}
\item Systems able to forecast $\geq M$ flares only.
\item Systems designed for exact 24, 48, and 72 hours of prediction, and also the next 24, 24-48, 48-72, and 72-96 hours.
\item Features, despite their origin (i.e., photospheric magnetic data, ARs photosferic features, among others)
\item Algorithms, despite their nature (i.e., SVMs, ensembles, decision trees, among others) 
\item Results from human predictions at forecast centers.
\end{itemize}

\begin{table*}
\centering
\caption{Literature state-of-the-art.}
\label{tab:results-comparison}
\resizebox{0.52\textwidth}{!}{
\begin{threeparttable}
\centering
\begin{tabular}{ccccccc}
\hline
\textit{Forecasting Time} & \textit{Authorship}      & \textit{Operationally-evaluated} & \textit{ACC} & \textit{TPR} & \textit{TNR} & \textit{TSS} \\ \hline
0-24           & GradientTreeBoosting\tnote{a}     & no                           & 0.91         & 0.93         & 0.88         & 0.81         \\
24-48          & GradientTreeBoosting\tnote{a}     & no                           & 0.92         & 0.94         & 0.88         & 0.82         \\
48-72          & GradientTreeBoosting\tnote{a}     & no                           & 0.94          & 0.97         & 0.89         & 0.86         \\
72-96          & GradientTreeBoosting\tnote{a}     & no                           & 0.94          & 0.98         & 0.89         & 0.86         \\
0-24         & \cite{Yang:2013}\tnote{b}       & no                           & 0.90         & 0.41         & 0.96         & 0.48         \\
0-48         & \cite{Yang:2013}\tnote{b}       & no                           & 0.86         & 0.43         & 0.95         & 0.53         \\
0-48         & \cite{Liu:2017a}\tnote{c}        & no                           & -            & 0.64         & 0.83         & 0.47            \\
0-24         & \cite{Liu:2017c}\tnote{d}   & no                             & 0.76         & 0.74         & 0.78         & 0.53         \\ 
0-48         & \cite{Liu:2017b}\tnote{e}   & no                             & 0.75         & 0.76         & 0.74         & 0.50         \\ 
0-48         & \cite{Li:2013}\tnote{f}         & no                           & 0.82         & 0.69         & 0.83         & 0.52         \\
0-48         & \cite{Li:2011}\tnote{g}         & no                           & 0.74         & 0.69         & 0.75         & 0.44           \\
0-48         & \cite{Huang:2013}\tnote{h}      & no                           & 0.72         & 0.72         & 0.71         & 0.71         \\
0-48         & \cite{Zhang:2011}\tnote{i}      & no                           & -            & 0.75         & -            & -            \\
0-48         & \cite{Yu:2009}\tnote{j}         & no                           & -            & 0.82         & 0.84         & 0.66         \\
0-48         & \cite{Yu:2010a}\tnote{k}        & no                           & 0.92         & 0.94         & 0.91         & 0.86         \\
0-48         & \cite{Yu:2010b}\tnote{l}        & no                           & -            & 0.85         & 0.87         & 0.72         \\
24         & \cite{Bobra:2015}\tnote{m}      & no                           & 0.92         & 0.83         & 0.92         & 0.76         \\
48         & \cite{Bobra:2015}\tnote{m}      & no                           & 0.94         & 0.86       & 0.94         & 0.81         \\
0-48         & \cite{Raboonik:2016}\tnote{n}   & no                             & 0.94         & 0.97         & 0.88         & 0.85         \\
0-24       & \cite{Muranushi:2015}\tnote{o}   & no                             & 0.7         & 0.85         & 0.67         & 0.52         \\
0-24       & \cite{Jonas:2018}\tnote{p}   & no                             & -         & -         & -         & 0.81         \\
0-48         & \cite{Huang:2010}\tnote{q}   & no                             & -         &  0.91         &  0.87         & 0.78         \\
0-24         & \cite{Huang:2018}\tnote{r}   & no                             & 0.81         &  0.85         &  0.81         & 0.66         \\
0-48         & \cite{Huang:2018}\tnote{r}   & no                             & 0.81         &  0.81         &  0.81         & 0.62         \\
0-24       & \cite{Sadykov:2017b}\tnote{s}   & no                             & 0.87         &  0.89         &  0.86         & 0.76         \\
0-24       & \cite{Nishizuka:2017}\tnote{t}  & no                              & 0.99         & 0.90         & 0.99         & 0.90  \\
\hline
0-24       & GradientTreeBoosting\tnote{a}     & yes                          & 0.71         & 0.75         & 0.70         & 0.46         \\
24-48      & GradientTreeBoosting\tnote{a}     & yes                          & 0.65         & 0.70         & 0.64         & 0.34         \\
48-72      & GradientTreeBoosting\tnote{a}     & yes                          & 0.70         & 0.75         & 0.69         & 0.44         \\
72-96      & GradientTreeBoosting\tnote{a}     & yes                          & 0.69         & 0.73         & 0.68         & 0.42         \\
0-24       & \cite{Bloomfield:2012}\tnote{u} & yes                          & 0.83         & 0.70         & 0.83        & 0.53         \\
0-24       & \cite{Kubo:2017}\tnote{v}& yes                  & 0.84         & 0.60         & 0.90        & 0.50 \\
0-48       & \cite{Devos:2014}\tnote{w}& yes                  & 0.88     & 0.37         & 0.97        & 0.34 \\
0-24       & \cite{Shin:2016}\tnote{x} & yes                  & -     & 0.61         & 0.76        & 0.37 \\
0-24       & \cite{Leka:2018}\tnote{y} & yes                  & 0.89     & 0.20         & 0.99        & 0.19 \\
24-48      & \cite{Leka:2018}\tnote{y} & yes                  & 0.87     & 0.03         & 1.00        & 0.03 \\
48-72      & \cite{Leka:2018}\tnote{y} & yes                  & 0.87     & 0.06         & 1.00        & 0.05 \\
0-24       & \cite{Nishizuka:2018}\tnote{z} & yes                  & 0.86     & 0.95         & 0.86        & 0.80 \\
0-24       & \cite{Hada-Muranushi:2016}\tnote{aa} & yes                  & 0.82     & 0.39         & 0.88        & 0.27 \\
0-24       & \cite{Crown:2012}\tnote{ab} & yes                  & 0.97     & 0.56         & 0.98        & 0.53 \\
0-24       & \cite{Anastasiadis:2017}\tnote{ac} & yes                  & -     & -         & -        & 0.25 \\
0-24       & \cite{McCloskey:2018}\tnote{ad} & yes                  & -     & -         & -        & 0.47 \\
0-24       & \cite{Falconer:2011,Falconer:2014}\tnote{ae} & yes                  & 0.95     & 0.31         & -        & 0.47 \\
0-24       & \cite{Falconer:2014}\tnote{af} & yes                  & 0.95     & 0.38         & -        & 0.49 \\
\bottomrule 
\end{tabular}

\begin{tablenotes}
\item[a] Our model.
\item[b] Results collected from Table 4. TPR treated as frequency of hits (FOH). TNR defined as frequency of correct nulls forecasts (FOCN)~\citep{Yang:2013}.
\item[c] Results collected from Tables 3 and 4. Authors did not inform the ACC. TSS calculated over TPR and TNR~\citep{Liu:2017a}.
\item[d] Results collected from Table 4~\citep{Liu:2017c}.
\item[e] Scores collected from Table 6. TSS calculated over TPR and TNR~\citep{Liu:2017b}.
\item[f] Results collected from Table 2. ACC is treated as CORR. Results refer to the w = 0 column. TSS calculated over TPR and TNR~\citep{Li:2013}.
\item[g] Results collected from Table 3. ACC is treated as correctness. Results refer to the KM-LVQ column. TSS calculated over TPR and TNR~\citep{Li:2011}.
\item[h] Scores computed over the confusion matrix of Table 3~\citep{Huang:2013}.
\item[i] Score gathered from Table III. Authors did not inform the TSS, TNR, and ACC. Results refer to the C4.5 column~\citep{Zhang:2011}.
\item[j] Scores gathered from Table 3. TSS computed over TPR and TNR. Results refer to the LVQ (w = 45) column~\citep{Yu:2009}. 
\item[k] Scores computed over the confusion matrix of Table 4. Results refer the MODWT\_DB2\_Red model~\citep{Yu:2010a}.
\item[l] Scores collected from Table 5. TSS computed over TPR and TNR. Results refer to the BN\_F column~\citep{Yu:2010b}.
\item[m] Results collected from Table 3~\citep{Bobra:2015}.
\item[n] Results collected from Table 3~\citep{Raboonik:2016}.
\item[o] Scores calculated over the confusion matrix of Figure 5. Results refer to a full-disk forecast~\citep{Muranushi:2015}.
\item[p] The TSS refers to the highest score of Figure 14 (24 h prediction task, features included: HMI and flare hist). Authors did not inform  ACC, TPR, and TNR~\citep{Jonas:2018}.
\item[q] Approximated scores from Figure 5 (in the graph, refer to the number of base prediction models equals 11). TSS calculated over TPR and TNR. Authors did not provide ACC~\citep{Huang:2010}.
\item[r] Scores calculated over the confusion matrix of Table 4~\citep{Huang:2018}.
\item[s] Scores computed from the TP, TN, FP, and FN values available at page 7~\citep{Sadykov:2017b}.
\item[t] Results computed over the confusion matrix of Table 3. Results refer to the k-NN model~\citep{Nishizuka:2017}.
\item[u] Scores collected from Table 4. TNR computed over TSS and TPR. Results refer to the optimum TSS entry~\citep{Bloomfield:2012}.
\item[v] Results collected from Table 4. TNR calculated over TSS and TPR~\citep{Kubo:2017}.
\item[w] Results collected from Table 3. TNR calculated over TSS and TPR~\citep{Devos:2014}.
\item[x] Scores collected from Tables 6 and 10. Results refer to the MLR1 model. TNR calculated over TPR and TSS~\citep{Shin:2016}.
\item[y] In Figure 13, authors provided full-disk and AR-by-AR scores, however, these results refer to the full-disk performance for a more concise reference~\citep{Leka:2018}.
\item[z] Scores calculated over the confusion matrix of Figure 5~\citep{Nishizuka:2018}.
\item[aa] Scores calculated over the confusion matrix of Table 5~\citep{Hada-Muranushi:2016}.
\item[ab] Scores calculated over the confusion matrix of Table 4~\citep{Crown:2012}.
\item[ac] Approximated TSS value from the graph of Figure 8 (TSS peak at a threshold of 0.15)~\citep{Anastasiadis:2017}.
\item[ad] TSS collected from the graph of Figure A.1 (p = 0.08 in the evolution line). Author did not inform TPR, TNR, and ACC~\citep{McCloskey:2018}.
\item[ae] Scores collected from the Table 2 of \cite{Falconer:2014} (Present MAG4 entry).
\item[af] Scores collected from the Table 2 of \cite{Falconer:2014} (Next MAG4 entry).
\end{tablenotes}

\end{threeparttable}

}
\end{table*}

\subsection{Not-operationally-evaluated Systems}
\label{sec:not-operationally-evaluated Systems}
Table~\ref{tab:results-comparison} shows the results of our case study with the gradient tree boosting models along with other papers. We divided this table into two parts: the upper-hand part represents not-operationally-evaluated models, while the lower-hand has the operationally-evaluated ones. 

\cite{Yang:2013}'s paper used photospheric AR vector magnetic data from the Solar Magnetic Field Telescope, located at the Huairou Solar Observing Station in China. Their models input features included but were not restricted to the mean planar magnetic shear angle, the vector magnetic field mean shear angle, and the mean absolute vertical current density. Then, \cite{Yang:2013} proposed a support vector classifier (SVC) using 10-fold cross-validation over the whole set of data. 

Since \cite{Yang:2013}'s paper only included magnetograms with ARs that were within \ang{30} from the solar disk center, we did classify their approach as a biased one. Nevertheless, they scored a TPR, TNR, ACC, and TSS score of 0.41, 0.96, 0.90, and 0.48, respectively, for 24 hours, and 0.43, 0.95, 0.86, and 0.53, for 48 hours. 

\cite{Liu:2017a}, in turn, proposed another case study that only included magnetograms with ARs within \ang{30} from the disk center. The authors used a mixed scenario of input features: on the one hand, they picked several magnetic field data, such as the neutral line length, the maximum horizontal gradient, and the singular points number from magnetograms taken by the Michelson Doppler Imager (MDI) \citep{Scherrer1995a} instrument aboard the Solar and Heliospheric Observatory (SOHO); on the other hand, they picked the \cite{McIntosh:1990} class of each AR in the sunspot catalog from the National Geophysical Data Center (NGDC).

\cite{Liu:2017a} then designed a multi-model integrated learner (MIM) by fitting several distinct base learners, such as neural networks, naïve classifiers, and SVMs. Base learners outputs were combined through linear sum, with weights adjusted by a genetic algorithm. As \cite{Yang:2013} did, \cite{Liu:2017a} also evaluated their MIM learner through 10-fold cross-validation, which made them to score TSS = 0.47. 

Similarly to \cite{Liu:2017a}'s methodology, \cite{Liu:2017b} used the same criterion to select MDI magnetograms for their dataset, i.\,e. they only included samples with ARs located within \ang{30} from the disk center. Instead of a multi-model learner, the authors used image-case-based reasoning to predict flares within 48 hours, which led them to score TSS = 0.5.

In turn, in \cite{Liu:2017c}'s paper, the focus was to evaluate the performance of a random forest algorithm to predict M- and X-class flares over Helioseismic and Magnetic Imager (HMI) magnetograms \citep{Bobra:2014} with a forecasting horizon of 24 hours. Using an imbalanced dataset with 24\% of positive samples, the authors chose downsample to treat this undesirable nature of their data. However, they did not reserve test sets at the beginning. Instead, they randomly downsampled 100 subsets until the positive and negative sample ratios were equal. 

Then, \cite{Liu:2017c} performed repeated 10-fold cross-validation in each downsampled subset, which ended up by showing a mean TNR of 0.78 and TPR of 0.74 (TSS = 0.53). Besides not properly reserving test sets, they also only picked magnetograms with ARs within \ang{70} of the Sun's central meridian. Therefore, their system was classified as not-operationally-evaluated.

As well as \cite{Liu:2017a}, \cite{Li:2013} also used 10-fold cross-validation to evaluate the performance of their model (multi-layer perceptron). Regarding 48 hours of prediction, the initial assembled dataset held samples spanning from April 1996 to December 2008. Besides, they also included features similar to ours (the sunspots area, magnetic classes, and x-ray fluxes) and  designed them through the sliding time window approach.

Because of the long records period, \cite{Li:2013}'s dataset naturally suffered from imbalanced class ratios. To tackle this undesirable issue, they proposed a $k$-means based undersample, further detailed in \cite{Li:2011}. However, they performed the under-sampling process before reserving the 10-fold cross-validation test sets, thus leading us to classify their results as having some bias, even though their neural network scored 0.69, 0.83, and 0.82 for TPR, TNR, and ACC, respectively. 

The unsupervised under-sampling process used by \cite{Li:2013} is detailed in \cite{Li:2011}'s paper, whose focus is also to propose a flare forecasting model, namely a learning vector quantization (LVQ) classifier. Regarding the resampling schema, \cite{Li:2011}'s methodology focused on dividing their dataset into positive and negative samples. The negative part was then inputted into a $k$-means algorithm with the $k$ value set to be the same number of flaring samples. 

Then, \cite{Li:2011} picked the clusters centroids and combined them with the positive samples before performing 10-fold cross-validation with their model. As \cite{Li:2013} suffered from some bias in their results for not reserving test data at the beginning, so did \cite{Li:2011}.

Despite \cite{Li:2011} and \cite{Li:2013}'s papers scored equal TPR results (0.69), their TNR differed: whereas the former scored 0.75, the latter equaled 0.83. This could be used to explain the difference of 0.08 between their TSS values, since this index can be interpreted as a direct response to TPR and TNR.

In addition to only including magnetograms with ARs near the disk center, the papers of \cite{Li:2013}, \cite{Li:2011}, \cite{Liu:2017a}, and \cite{Liu:2017b} also fall in the third criterion to be classified as not-operationally-evaluated. Thus, they designed the negative class differently from our models. While we consider the M- or X-class flares existence to flag a positive example and C-class events or the flares nonexistence to represent a negative one, those authors only consider the existence of C-class flares as the negative example. 

Another paper that employed the same design to negative examples and also only picked magnetograms with ARs within \ang{30} from the disk center was proposed by \cite{Huang:2013}. Using features based on data from highly stressed longitudinal magnetic fields taken from MDI magnetograms, the authors designed a single decision tree to forecast $\geq M1.0$ flares 48 hours ahead.

Similarly to \cite{Huang:2013}, \cite{Zhang:2011} also only picked magnetograms with ARs linked to $\geq C1.0$ flares to be in their dataset and included samples within \ang{30} from the disk center. Using features such as the magnetic field and texture distribution, and the largest sunspot group fractal dimensional, the authors designed a single C4.5 decision tree to forecast $\geq M1.0$ flares 48 hours ahead. The performance of their system was evaluated by 10-fold cross-validation, with only a reported TPR of 0.75. 

\cite{Yu:2009}, \cite{Yu:2010a}, and \cite{Yu:2010b} proposed other papers that also only picked magnetograms with ARs linked to $\geq C1.0$ flares to be in their datasets and discarded samples far from the disk center. The authors calculated the maximum horizontal gradient, the neutral line length, and the singular points number over ARs taken from MDI magnetograms and used them as features for all papers. Besides, all of them used the sliding time window schema to represent data evolution. However, the authors differed regarding the algorithms used: in \cite{Yu:2009}, a C4.5 and a LVQ model were used; in \cite{Yu:2010b}, the model was a Bayesian network; and in \cite{Yu:2010a}, they used the C4.5 tree again. 
Following this trend, \cite{Huang:2010} pointed out to use the same data selection criteria for picking ARs samples as \cite{Yu:2009}, which also made them to be classified as having some bias.

\cite{Bobra:2015}'s paper, in turn, treated positive and negative classes differently from \cite{Yu:2009}, \cite{Yu:2010a}, \cite{Yu:2010b}, \cite{Li:2013}, \cite{Li:2011}, \cite{Liu:2017a}, \cite{Liu:2017b}, and \cite{Huang:2013}. They used $\geq M1.0$ x-ray measures to flag positive samples while defining C-class flares and the events nonexistence as the negative cases. Their dataset included 303 positive samples and 5000 randomly selected negative examples taken from HMI magnetograms, which represented a imbalanced case of class ratios.

To cope with their imbalanced data nature, \cite{Bobra:2015} used the cost function of a SVM classifier instead of a resampling schema. Hence, they assigned different weights to their positive and negative classes to guarantee the classifier did not give much emphasis to the negative samples. To validate their model, they used repeated random subsampling, thus splitting their data into training (70\%) and test (30\%) sets at each iteration.

In fact, what \cite{Bobra:2015} did was to verify how their TSS changed in the test sets according to the weights variation between positive and negative classes. Then, they picked the optimal weight value so they could in their own words maximize the TSS in the test sets.

However, \cite{Bobra:2015}'s rationale for adjusting the TSS came out with the bias of producing tailored results regarding the test sets used, i.e., their proposal fell in the first criterion of not-operationally-evaluated systems. Besides tayloring the results for their test sets, \cite{Bobra:2015} also only included AR samples within \ang{68} of the disk center. In addition, because their dataset only held about 300 positive examples, this could affect their model generalization skill in real operational settings.  

\cite{Jonas:2018} designed a time series dataset to forecast the M- and X-class flares existence within the next 24 hours along with linear classifiers. Their dataset referred to the same HMI data period as \cite{Bobra:2015}. To evaluate their models, they tried different combinations of features using repeated random subsampling schema, splitting samples into training (80\%) and testing (20\%) data.  

\cite{Jonas:2018}'s aim was to report the best features subset, i.e., the one that best increased the TSS over the test sets. Since the authors used test data for decision-making instead of to play the role of unseen samples, we classified their results as not-operationally-evaluated. Test data samples were used as validation data in fact \citep{Hastie:2009}. 

\cite{Raboonik:2016} proposed another model that used a reduced dataset during training as \cite{Bobra:2015} did. Despite authors achieved high scores with their SVM for predicting flares 48 hours ahead (TPR = 0.97, TNR = 0.88, ACC = 0.94, and TSS = 0.85), they included in their dataset only 85 positive and 208 negative class samples. As previously commented, reduced datasets weaken the interpretation of how systems behave in an operational sense, since the number of examples is too small and models probably would not generalize well if put into an operational setting \citep{Pyle:1999}. 

Besides proposing the UFCORIN engine for fitting regression predictors over time series, \cite{Muranushi:2015} also carried out a case study to prove their pipeline effectiveness. Thus, they assembled a time series involving several features calculated from HMI magnetograms. However, the authors only picked AR samples within \ang{69} of the disk center. Other papers that also included ARs near the center are \cite{Huang:2018} (\ang{30}) and \cite{Sadykov:2017b} (\ang{68}).

\cite{Nishizuka:2017}, in turn, were the first authors that included magnetograms with ARs beyond the limb at their models. From 2010 to 2015, they calculated about 60 features from full-disk HMI magnetograms, including those proposed by \cite{Bobra:2015} and several others related to UV brightening and flare history. Then, they compared the performance of some machine learning methods, such as k-NN, SVM, and extremely randomized trees, through repeated random subsampling based on a data split into training (70\%) and test (30\%) sets to find the best algorithm regarding the mean TSS. 

However, \cite{Nishizuka:2017}'s rationale for picking the best model relying on the mean TSS of  test sets was similar to \cite{Bobra:2014}'s and \cite{Jonas:2018}'s, who also used the test sets for decision-making. Thus, we  classified their paper as not-operationally-evaluated for not reserving real unseen data and keeping them aside from the evaluation process while fitting their models. Despite, their k-NN model had an ACC, TPR, TNR, and TSS equaling 0.99, 0.90, 0.99, and 0.90, respectively.

\subsection{Operationally-evaluated Systems}
\label{sec:operationally-evaluated Systems}
Conversely to the previous papers that had some bias during their performance evaluation, \cite{Bloomfield:2012}'s paper presented a methodology that was free from this undesirable issue. Using data from the sunspot region summaries provided by the NOAA/SWPC, they used Poisson statistics to calculate the flaring probabilities of each \cite{McIntosh:1990} class. To evaluate their model, they split their dataset into years: the period between 1988 and 1996 was used as the training set while 1996 to 2010 was their test set. Considering 24 hours of $\geq M1.0$ flares prediction, they scored 0.70, 0.83, 0.83, and 0.53, respectively, for TPR, TNR, ACC, and TSS.

Similarly to \cite{Bloomfield:2012}, \cite{McCloskey:2018} also used Poisson statistics over the \cite{McIntosh:1990} classes. However, they used the evolution of such classes in sunspot groups instead of focusing on static point-in-time observations. Their method training period was the years between 1988 and 1996, with more recent samples from 1996 to 2008 used as test data.

\cite{Shin:2016}, in turn, designed an AR-by-AR prediction schema using multiple linear regression. For the ARs in the NOAA/SWPC sunspot region summaries, they calculated the weighted mean flare rate of each \cite{McIntosh:1990} class and magnetic configuration, as well as the previous day weighted total flare flux. Then, they randomly sampled several C-, M-, and X-class events from January 1996 to December 2004 for training their model while using all flare events from January 2005 to November 2013 for testing.

Although we previously classified \cite{Muranushi:2015}'s results as not-operationally-evaluated for only including magnetograms with ARs near the solar disk center in their dataset, there was an attempt to deploy an operational system designed by their UFCORIN engine recently, as described in \cite{Hada-Muranushi:2016}. Instead of a regression model, authors rewrote the UFCORIN to fit a long-short term memory neural network (LSTM) upon wavelet features calculated over HMI images.

\cite{Falconer:2011} presented another system that was already deployed into an operational setting, namely the Magnetogram Forecast Forecasting Tool (MAG4). This tool was responsible for monitoring and forecasting astronauts radiation exposure levels by predicting M‐ and X‐class flares, coronal mass ejections, and solar energetic particle events. MAG4 was built over a dataset of 40.000 magnetograms from 1.300 active regions taken by the SOHO/MDI instrument.

Although MAG4 was used within a deployment environment, it was designed to include only ARs within \ang{30} of the solar disk center. For predictions beyond this radius, the tool warned reduced performance. Although operationally ready, there is uncertainty about the reported scores (ACC = 0.95, TPR = 0.31, and TSS = 0.47).

More recently, using the same data as \cite{Falconer:2011}, there was an attempt to improve MAG4 performance by using previous flare history with features that characterize free-energy in ARs \citep{Falconer:2014}. This approach increased the tool original TPR and TSS scores by 0.07 and 0.02, respectively.

Finally, using the same data period as \cite{Nishizuka:2017} but splitting the period from 2010 to 2014 for training and reserving 2015 for testing, \cite{Nishizuka:2018} designed a convolutional neural network (CNN) to predict $\geq M$ class flares within 24 hours.

\subsection{Space Weather Prediction Centers Forecasts}
The last class of papers included in Table~\ref{tab:results-comparison} comprehends human-based forecasts from space weather prediction centers. Here, we included three proposals found regarding $\geq M$ class flares forecasting, namely \cite{Crown:2012}, \cite{Kubo:2017}, and \cite{Devos:2014}. 

At NOAA/SWPC, from which we based our input features and forecasting horizons, the TSS was 0.49 for predicting M-class flares and above in the next 24 hours as pointed out by~\cite{Crown:2012} in the solar cycle 23 (May 1996 to December 2008). In turn, the ACC, TPR, and TNR were 0.97, 0.56, and 0.98, respectively, with the FAR = 0.57.

The second paper we cited is from the SWPC of Japan National Institute of Information and Communications Technology (NICT), where human forecasters scored TSS = 0.5 to predict $\geq M$ class flares within 24 hours. In turn, their ACC and TPR were, respectively, 0.84 and 0.60 in the period between 2000 and 2015. However, their FAR = 0.42~\citep{Kubo:2017}.

Finally, at the Solar Influences Data Center (SIDC) of the Royal Observatory of Belgium (ROB), the TSS was 0.34 for predicting flares within the next 48 hours during the period from June 2004 to December 2012 \citep{Devos:2014}. Besides, their human-generated forecasts did score TPR = 0.37 and TNR = 0.97, respectively. Their FAR, in turn, was only 0.35.

\subsection{Remarks from the Literature Analysis}

As we can see, papers mentioned earlier differed a lot concerning their forecasting horizons, features, and algorithms, even though their main goal was to predict at least M-class events. Therefore, results commented in the next subsections should not be considered direct comparisons of scores between different systems. Instead, we shall analyze systems according to some identified interplay effects involving accuracy, recall, precision, and the number of false alarms, seeking over-fitted systems.

\subsubsection{Remarks from the Not-operationally-evaluated Systems Analysis}

Except for the papers by \cite{Zhang:2011} and \cite{Jonas:2018}, whose authors did not provide any other metric than TSS or TPR, we managed to distinguish models within this section into groups comprehending with or without over-fitting.

Regarding models forecasting events in the next 24 hours, the group without over-fitting included our model, \cite{Liu:2017c, Muranushi:2015, Huang:2018, Sadykov:2017b}; and \cite{Nishizuka:2017}. As they scored high ACCs (\big[0.70,0.99\big]), TPRs (\big[0.74,0.93\big]), and TNRs (\big[0.67,0.99\big]), we could not note any over-fitted performance here. 

However, despite having achieved high TPRs (\big[0.85,0.89\big]), papers by \cite{Muranushi:2015, Huang:2018}; and \cite{Sadykov:2017b} have output many false alarms in their predictions, noticeably, FAR = 0.63, 0.90, and 0.76, respectively. The precision of their classifiers might have caused those mispredictions.

Directly related to TPR, the precision -- as defined by \cite{Zaki:2013} -- represents the accuracy for predicting positive events. While the former represents the number of positive events correctly predicted, the latter accounts for the skill to predict positive events.

Often, there will be a trade-off between recall and precision. For instance, as argued by \cite{Zaki:2013}, it would be rather easy to score TPR = 1 by merely predicting all testing samples as positive. However, the precision would be rather low, thus highly increasing the number of false alarms.

Conversely, we could highly increase the precision provided that we predict only a few testing samples as positive (for instance, the samples about which our model has the most confidence). In this sense, the recall would be low, and thus the number of false alarms would also be low.

In fact, what we argue here is that by trying to boost TPR, the authors outputting high FAR numbers may have harmed the precision of their systems, thus increasing the number of predicted false alarms.

If, on the one hand, we had several examples that did not over-fit to a specific class when forecasting within the next 24 hours, on the other hand, we had only one model composing the group of classifiers that over-fitted, namely the \cite{Yang:2013}'s classifier. The authors have achieved both ACC (0.90) and TNR (0.96) at high levels simultaneously. Their TPR, in turn, only scored 0.41. Thus, we could suggest over-fitting in favor of the negative class.

Regarding models forecasting events in the next 48 hours, the group of classifiers without over-fitting included \cite{Liu:2017a, Liu:2017b, Li:2013, Li:2011, Huang:2013, Yu:2009, Yu:2010a, Yu:2010b, Raboonik:2016, Huang:2010}; and \cite{Huang:2018}. For having scored high ACCs (\big[0.72,0.94\big]), along with TPRs (\big[0.64,0.97\big]) and TNRs (\big[0.71,0.91\big]), also at high levels, we could also not note any over-fitted class here. 

However, as we had some models observing false alarms for predictions in the next 24 hours, so did we for the next 48. Despite scoring high TPRs (\big[0.72,0.94\big]), papers by \cite{Huang:2013, Yu:2010a, Yu:2010b}; and \cite{Huang:2018} experienced false alarms to some extent in their predictions, namely, FAR = 0.70, 0.29, 0.28, and 0.84, respectively. Those mispredictions might have been caused again by the precision of their classifiers, that is, the precision-recall trade-off, which directly affected FAR.

Finally, the remainder of not-operationally-evaluated classifiers included those designed for other forecasting horizons, such as our models for 24-48, 48-72, and 72-96 hours, and \cite{Bobra:2015}'s, with forecasts exactly after 24 and 48 hours. For those classifiers, we did not need any distinguishment since all of them did not over-fit.

However, \cite{Bobra:2015}'s classifiers have output false alarms to some extent, probably because of the precision-recall trade-off. Despite scoring TPR = 0.83 (exact 24 hours) and 0.86 (exact 48 hours) and not providing their FAR values, \cite{Bobra:2015} reported harmed precision scores, noticeably PPV = 0.41 (exact 24 hours) and 0.50 (exact 48 hours).

\subsubsection{Remarks from the Operationally-evaluated Systems Analysis}

For predictions in the next 24 hours, the group of models that over-fitted to a specific class included \cite{Leka:2018}, \cite{Hada-Muranushi:2016}; and both models by \cite{Falconer:2014}. On the one hand, as \cite{Hada-Muranushi:2016} and \cite{Leka:2018}'s TPRs varied over {\big[0.20,0.39\big]} and TNRs {\big[0.88,0.99\big]}, those systems could not generalize positive samples well, that is, their ACCs corroborate the existence of over-fitting ({\big[0.82,0.89\big]}). Besides, we could also note a high number of false alarms with \cite{Hada-Muranushi:2016} (FAR = 0.68), which directly suggests low precision.

On the other hand, \cite{Falconer:2014} only provided their ACCs (both models equaled 0.95) and recall scores (TPR = 0.31 and 0.38). In this sense, we could suggest here high TNRs since they scored high ACCs. In fact, what we argue is that this scenario indirectly suggests over-fitting, since ACC is highly influenced by over-performing with a specific class and they have low TPRs. Besides over-fitting, their models also had their precision skills harmed because of the high numbers of false alarms (FAR = 0.5 and 0.48).

In turn, the group of models without over-fitting for forecasting in the next 24 hours included our model, \cite{Shin:2016}, \cite{Bloomfield:2012}; and \cite{Nishizuka:2018}. Given the ACCs in a high interval ({\big[0.71,0.86\big]}) with both TPRs and TNRs simultaneously at high levels ({\big[0.61,0.95\big]} and {\big[0.70,0.86\big]}, respectively), we could not suggest over-fitting. However, despite this absence, some models did have their precision skills harmed because of the high numbers of false alarms, noticeably \cite{Shin:2016} (FAR = 0.78), \cite{Nishizuka:2018} (FAR = 0.82), and \cite{Bloomfield:2012} (FAR = 0.82).

For longer forecasting horizons, namely 24-48 (our model and \cite{Leka:2018}'s), 48-72 (our model and \cite{Leka:2018}'s), and 72-96 hours (our model), we could only observe over-fitting with \cite{Leka:2018}'s classifiers. Both scored TNR = 1 and ACC = 0.87, along with low TPRs (0.03 and 0.06). Finally, regarding the remainder of papers, classifiers by \cite{Anastasiadis:2017} and \cite{McCloskey:2018} did not provide any other metric than TSS.

\section{Conclusions}\label{sec:conclusions}

Searching for accurate models for solar flares forecasting and looking for an attempt to create some standardization when designing them, we proposed a framework to deal with the most common aspects coped by literature while designing such systems. Our framework includes feature selection, hyper-parameters fine-tuning, evaluation under operational settings, imbalanced dataset resampling and selection between distinct models.

Therefore, this paper presented a framework to design, train, and evaluate flare forecasting systems with flexibility and performance. Framework flexible aspects include:

\begin{enumerate}[(i)]
\item \textit{Event magnitude adjustment}: we set the prediction to be over $\geq M$ class flares in our case study. However, depending on the features set used, the event magnitude could be changed to other thresholds or even to allow the forecasting of specific flare classes.
\item \textit{Inner scores optimization}: TSS is not the only option available to optimize between inner processes. Other available scores would be AUC, precision, positive recall, f1-score, etc.
\item \textit{Prediction vs. recall threshold shifting}: although we did not set up a custom prediction threshold while distinguishing events as positives or negatives, this could be done by adjusting the output of algorithms -- in our case, we set our models thresholds to the default level, 0.5.
\item \textit{Custom algorithms}: provided that new algorithms are supplied with their corresponding hyper-parameter grids, new forecasting models would be designed, trained, and validated. Thus, we did not restrict our framework to the case study algorithms, since they are treated as ``black boxes''.
\item \textit{Forecasting other events than solar flares}: since models output can be changed with ease (i.e., by providing other target features), the framework can be used to forecast other events, such as coronal mass ejections, and mass and speed of solar energetic particles, which make it have a broader application in solar weather research.
\item \textit{Resampling method}: we did not restrict our framework to SMOTE-ENN. In this sense, the algorithm for coping with imbalanced datasets would be changed with ease, i.e., other feasible techniques include random over- or under-sampling, a simple SMOTE, or even other advanced resampling algorithms.
\item \textit{Feature selection}: although we did propose a univariate feature selection method provided with the F-score, other algorithms could be used in the framework pipeline, such as the Pearson correlation analysis.
\item \textit{Search for hyper-parameters}: instead of randomly searching for hyper-parameters, searching the whole parameters grid is also possible.
\item \textit{Criteria for supporting the inner cross-cutting decisions}: for instance, to pick the representative features, intersecting between those with the highest TSSs was solely our case study. We could have considered other approaches for picking the representative set, i.e., pick the features set with the highest score among all validation sets.
\item \textit{Custom pipeline flow}: since our framework treats its inner processes as blocks with well-defined input and output elements, several other machine-learning techniques other than model selection, feature selection, hyper-parameters optimization, and data resampling could be used with ease (for instance, processes to adjust classifiers cut-off points or to cope with cost-sensitive learning).

\end{enumerate}

To validate our framework, we assembled a dataset based on NOAA/SWPC daily aggregated data from the Sun's behaviors, which comprehended several distinct predictive features, including the sunspot number and area, radio flux, x-ray background flux, and the most common ARs magnetic classes. We aimed at forecasting major solar flares (M class and above) up to 4 days ahead, namely the next 24, 24-48, 48-72, and 72-96 hours ahead. We also designed our input predictive features to be always observed five days before the forecasting periods.

Concerning the framework effectiveness, the gradient tree boosting schemas designed under our pipeline increased their recall scores over simulated operational scenarios by 0.40 (the next 24 and 24-48 hours), 0.53 (48-72 hours), and 0.57 (72-96 hours) while keeping their AUCs at high levels (those results refer to increases over baseline predictions). Besides, we could also observe that our models changed from over-fitted systems that used to favor the original negative class to versions that could forecast positive and negative classes without any preference. 

In turn, the TSS of models increased by 0.16, 0.8, 0.25, and 0.24 over our baseline level for the next 24, 24-48, 48-72, and 72-96 hours, respectively. Those increments were not higher since we had TNR considerably decreased by the chosen hybrid resampling method.

We also argue that our models could be used along with hybrid prediction schemas, such as the one employed at NOAA/SWPC, which has predictions at first calculated by an expert system and then forecasts are confirmed/adjusted by experts. Also, such as the one at MOSWOC/SWPC, which has their first forecast estimates made by Poisson statistics and then experts confirm/adjust them. This is needed to mitigate some climatology effects that may influence predictions made over extended validity periods as ours, i.e., how ARs evolve while transversing the disk or which ARs may be leaving or returning to the disk in the next few days \citep{Murray:2017}.

As future research, we plan to carry out other experiments with the framework focusing on refining its inner techniques to improve the output model performance and deploy it into a real operational setting. Since we based our training dataset mainly upon DSD data, we can think of an operational system that follows NOAA/SWPC's periodicity to update this data product. 

According to NOAA/SWPC's readme file\textsuperscript{\ref{swpc-readme}}, DSD data are always updated at 02:25, 08:25, 14:25, and 20:25 UT. Thus, our deployment environment is expected to run with a cadence of four times a day. At each run, the system shall collect the last five days of records available in DSD, integrate with SRS data, and then forecast the existence of $M+X$ flares up to four days ahead.

\section*{Acknowledgements}

This study was financed in part by the Coordenação de Aperfeiçoamento de Pessoal de Nível Superior (CAPES), Brazil -- Finance Code 001. Also, we thank: (i) Federal Institute of Education, Science and Technology of Rio Grande do Sul (IFRS) -- Campus Feliz, for the cooperation with this research; (ii) NOAA/SWPC, for the data; (iii) the reviewer, for the valuable comments and suggestions; and (iv) Espa\c{c}o da Escrita, Pr\'o-Reitoria de Pesquisa, UNICAMP, for the language services provided.



\balance
\bibliographystyle{mnras}
\bibliography{library.bib} 

\begin{thebibliography}{}
\makeatletter
\relax
\def\mn@urlcharsother{\let\do\@makeother \do\$\do\&\do\#\do\^\do\_\do\%\do\~}
\def\mn@doi{\begingroup\mn@urlcharsother \@ifnextchar [ {\mn@doi@}
  {\mn@doi@[]}}
\def\mn@doi@[#1]#2{\def\@tempa{#1}\ifx\@tempa\@empty \href
  {http://dx.doi.org/#2} {doi:#2}\else \href {http://dx.doi.org/#2} {#1}\fi
  \endgroup}
\def\mn@eprint#1#2{\mn@eprint@#1:#2::\@nil}
\def\mn@eprint@arXiv#1{\href {http://arxiv.org/abs/#1} {{\tt arXiv:#1}}}
\def\mn@eprint@dblp#1{\href {http://dblp.uni-trier.de/rec/bibtex/#1.xml}
  {dblp:#1}}
\def\mn@eprint@#1:#2:#3:#4\@nil{\def\@tempa {#1}\def\@tempb {#2}\def\@tempc
  {#3}\ifx \@tempc \@empty \let \@tempc \@tempb \let \@tempb \@tempa \fi \ifx
  \@tempb \@empty \def\@tempb {arXiv}\fi \@ifundefined
  {mn@eprint@\@tempb}{\@tempb:\@tempc}{\expandafter \expandafter \csname
  mn@eprint@\@tempb\endcsname \expandafter{\@tempc}}}

\bibitem[\protect\citeauthoryear{Ahmed, Qahwaji, Colak, Higgins, Gallagher  \&
  Bloomfield}{Ahmed et~al.}{2013}]{Ahmed:2013}
Ahmed O.~W.,  Qahwaji R.,  Colak T.,  Higgins P.~A.,  Gallagher P.~T.,
  Bloomfield D.~S.,  2013, \mn@doi [Solar Physics] {10.1007/s11207-011-9896-1},
  283, 157

\bibitem[\protect\citeauthoryear{Al-Ghraibah, Boucheron  \&
  McAteer}{Al-Ghraibah et~al.}{2015}]{Al-Ghraibah:2015a}
Al-Ghraibah A.,  Boucheron L.~E.,   McAteer R. T.~J.,  2015, \mn@doi [Astronomy
  {\&} Astrophysics] {10.1051/0004-6361/201525978}, 579, A64

\bibitem[\protect\citeauthoryear{Anastasiadis, Papaioannou, Sandberg,
  Georgoulis, Tziotziou, Kouloumvakos  \& Jiggens}{Anastasiadis
  et~al.}{2017}]{Anastasiadis:2017}
Anastasiadis A.,  Papaioannou A.,  Sandberg I.,  Georgoulis M.,  Tziotziou K.,
  Kouloumvakos A.,   Jiggens P.,  2017, \mn@doi [Solar Physics]
  {10.1007/s11207-017-1163-7}, 292, 21pp

\bibitem[\protect\citeauthoryear{Barnes \& Leka}{Barnes \&
  Leka}{2008}]{Barnes:2008}
Barnes G.,  Leka K.~D.,  2008, \mn@doi [The Astrophysical Journal]
  {10.1086/595550}, 688, L107

\bibitem[\protect\citeauthoryear{Barnes, Leka, Schumer  \& Della-Rose}{Barnes
  et~al.}{2007}]{Barnes:2007}
Barnes G.,  Leka K.~D.,  Schumer E.~A.,   Della-Rose D.~J.,  2007, \mn@doi
  [Space Weather] {10.1029/2007SW000317}, 5, 9pp

\bibitem[\protect\citeauthoryear{Barnes et~al.,}{Barnes
  et~al.}{2016}]{Barnes:2016}
Barnes G.,  et~al., 2016, \mn@doi [The Astrophysical Journal]
  {10.3847/0004-637X/829/2/89}, 829, 32pp

\bibitem[\protect\citeauthoryear{Batista, Prati  \& Monard}{Batista
  et~al.}{2004}]{Batista:2004}
Batista G. E. A. P.~A.,  Prati R.~C.,   Monard M.~C.,  2004, Sigkdd
  Explorations, 6, 20

\bibitem[\protect\citeauthoryear{Benvenuto, Piana, Campi  \& Massone}{Benvenuto
  et~al.}{2018}]{Benvenuto:2018}
Benvenuto F.,  Piana M.,  Campi C.,   Massone A.~M.,  2018, \mn@doi [The
  Astrophysical Journal] {10.3847/1538-4357/aaa23c}, 853, 9pp

\bibitem[\protect\citeauthoryear{Bergstra \& Bengio}{Bergstra \&
  Bengio}{2012}]{Bergstra:2012}
Bergstra J.,  Bengio Y.,  2012, Journal of Machine Learning Research, 13, 281

\bibitem[\protect\citeauthoryear{Bloomfield, Higgins, McAteer  \&
  Gallagher}{Bloomfield et~al.}{2012}]{Bloomfield:2012}
Bloomfield D.~S.,  Higgins P.~A.,  McAteer R. T.~J.,   Gallagher P.~T.,  2012,
  \mn@doi [The Astrophysical Journal] {10.1088/2041-8205/747/2/L41}, 747, L41

\bibitem[\protect\citeauthoryear{Bobra \& Couvidat}{Bobra \&
  Couvidat}{2015}]{Bobra:2015}
Bobra M.~G.,  Couvidat S.,  2015, \mn@doi [The Astrophysical Journal]
  {10.1088/0004-637X/798/2/135}, 798, 135

\bibitem[\protect\citeauthoryear{Bobra, Sun, Hoeksema, Turmon, Liu, Hayashi,
  Barnes  \& Leka}{Bobra et~al.}{2014}]{Bobra:2014}
Bobra M.~G.,  Sun X.,  Hoeksema J.~T.,  Turmon M.,  Liu Y.,  Hayashi K.,
  Barnes G.,   Leka K.~D.,  2014, \mn@doi [Solar Physics]
  {10.1007/s11207-014-0529-3}, 289, 3549

\bibitem[\protect\citeauthoryear{Breiman, Friedman, Olshen  \& Stone}{Breiman
  et~al.}{1984}]{Breiman:1984}
Breiman L.,  Friedman J.,  Olshen R.,   Stone C.,  1984, {Classification and
  Regression Trees}, 1st edn.
Wadsworth, Monterey, CA

\bibitem[\protect\citeauthoryear{Canfield}{Canfield}{2001}]{Canfield:2001}
Canfield R.~C.,  2001, in , Encyclopedia of the history of astronomy and
  astrophysics.
Institute of Physics Publishing, Bristol, UK, pp~1--6

\bibitem[\protect\citeauthoryear{Chang \& Lin}{Chang \& Lin}{2008}]{Chang:2008}
Chang Y.-W.,  Lin C.-J.,  2008, in JMLR: Workshop and Conference Proceedings.
  WCCI2008 Workshop on Causality. pp 53--64

\bibitem[\protect\citeauthoryear{Chawla, Bowyer, Hall  \& Kegelmeyer}{Chawla
  et~al.}{2002}]{Chawla:2002}
Chawla N.~V.,  Bowyer K.~W.,  Hall L.~O.,   Kegelmeyer W.~P.,  2002, Journal of
  Artificial Intelligence Research, 16, 321

\bibitem[\protect\citeauthoryear{Claesen \& {De Moor}}{Claesen \& {De
  Moor}}{2015}]{Claesen:2015}
Claesen M.,  {De Moor} B.,  2015, CoRR, abs/1502.0

\bibitem[\protect\citeauthoryear{Colak \& Qahwaji}{Colak \&
  Qahwaji}{2007}]{Colak:2007}
Colak T.,  Qahwaji R.,  2007, in , Vol.~39, Soft Computing in Industrial
  Applications.
Springer Berlin Heidelberg, Berlin, Heidelberg, pp 316--324,
  \mn@doi{10.1007/978-3-540-70706-6_29}, \url
  {http://link.springer.com/10.1007/978-3-540-70706-6{\_}29}

\bibitem[\protect\citeauthoryear{Colak \& Qahwaji}{Colak \&
  Qahwaji}{2009}]{Colak:2009}
Colak T.,  Qahwaji R.,  2009, \mn@doi [Space Weather] {10.1029/2008SW000401},
  7, 1

\bibitem[\protect\citeauthoryear{Crown}{Crown}{2012}]{Crown:2012}
Crown M.~D.,  2012, \mn@doi [Space Weather] {10.1029/2011SW000760}, 10, 4pp

\bibitem[\protect\citeauthoryear{Devos, Verbeeck  \& Robbrecht}{Devos
  et~al.}{2014}]{Devos:2014}
Devos A.,  Verbeeck C.,   Robbrecht E.,  2014, \mn@doi [Journal of Space
  Weather and Space Climate] {10.1051/swsc/2014025}, 4, A29

\bibitem[\protect\citeauthoryear{Domijan, Bloomfield  \& Piti{\'{e}}}{Domijan
  et~al.}{2019}]{Domijan:2019}
Domijan K.,  Bloomfield D.~S.,   Piti{\'{e}} F.,  2019, \mn@doi [Solar Physics]
  {10.1007/s11207-018-1392-4}, 294

\bibitem[\protect\citeauthoryear{Falconer, Barghouty, Khazanov  \&
  Moore}{Falconer et~al.}{2011}]{Falconer:2011}
Falconer D.,  Barghouty A.~F.,  Khazanov I.,   Moore R.,  2011, \mn@doi [Space
  Weather] {10.1029/2009SW000537}, 9, 12pp

\bibitem[\protect\citeauthoryear{Falconer, Moore, Barghouty  \&
  Khazanov}{Falconer et~al.}{2014}]{Falconer:2014}
Falconer D.~A.,  Moore R.~L.,  Barghouty A.~F.,   Khazanov I.,  2014, \mn@doi
  [Space Weather] {10.1002/2013SW001024}, 12, 306

\bibitem[\protect\citeauthoryear{Gallagher, Moon  \& Wang}{Gallagher
  et~al.}{2002}]{Gallagher:2002}
Gallagher P.,  Moon Y.-J.,   Wang H.,  2002, \mn@doi [Solar Physics]
  {10.1023/A:1020950221179}, 209, 171

\bibitem[\protect\citeauthoryear{Gower}{Gower}{1971}]{Gower:1971}
Gower J.~C.,  1971, \mn@doi [Biometrics] {10.2307/2528823}, 27, 857

\bibitem[\protect\citeauthoryear{Guerra, Pulkkinen  \& Uritsky}{Guerra
  et~al.}{2015}]{Guerra:2015}
Guerra J.~A.,  Pulkkinen A.,   Uritsky V.~M.,  2015, \mn@doi [Space Weather]
  {10.1002/2015SW001195}, 13, 626

\bibitem[\protect\citeauthoryear{Guyon \& Elisseeff}{Guyon \&
  Elisseeff}{2003}]{Guyon:2003}
Guyon I.,  Elisseeff A.,  2003, Journal of Machine Learning Research, 3, 1157

\bibitem[\protect\citeauthoryear{Hada-Muranushi, Muranushi, Asai, Okanohara,
  Raymond, Watanabe, Nemoto  \& Shibata}{Hada-Muranushi
  et~al.}{2016}]{Hada-Muranushi:2016}
Hada-Muranushi Y.,  Muranushi T.,  Asai A.,  Okanohara D.,  Raymond R.,
  Watanabe G.,  Nemoto S.,   Shibata K.,  2016, p.~6pp

\bibitem[\protect\citeauthoryear{Hale, Ellerman, Nicholson  \& Joy}{Hale
  et~al.}{1919}]{Hale:1919}
Hale G.~E.,  Ellerman F.,  Nicholson S.~B.,   Joy A.~H.,  1919, Astrophysical
  Journal, 49, 153

\bibitem[\protect\citeauthoryear{Han \& Kamber}{Han \& Kamber}{2006}]{Han:2006}
Han J.,  Kamber M.,  2006, {Data Mining: Concepts and Techniques}, 2 edn.
Morgan Kaufmann Publishers, San Francisco, USA

\bibitem[\protect\citeauthoryear{Hastie, Tibshirani  \& Friedman}{Hastie
  et~al.}{2009}]{Hastie:2009}
Hastie T.,  Tibshirani R.,   Friedman J.,  2009, {The Elements of Statistical
  Learning - Data Mining, Inference, and Prediction}, 2nd edn.
Springer, New York, NY, USA

\bibitem[\protect\citeauthoryear{Huang \& Wang}{Huang \&
  Wang}{2013}]{Huang:2013}
Huang X.,  Wang H.-N.,  2013, \mn@doi [Research in Astronomy and Astrophysics]
  {10.1088/1674-4527/13/3/010}, 13, 351

\bibitem[\protect\citeauthoryear{Huang, Yu, Hu, Wang  \& Cui}{Huang
  et~al.}{2010}]{Huang:2010}
Huang X.,  Yu D.,  Hu Q.,  Wang H.,   Cui Y.,  2010, \mn@doi [Solar Physics]
  {10.1007/s11207-010-9542-3}, 263, 175

\bibitem[\protect\citeauthoryear{Huang, Wang, Xu, Liu, Li  \& Dai}{Huang
  et~al.}{2018}]{Huang:2018}
Huang X.,  Wang H.,  Xu L.,  Liu J.,  Li R.,   Dai X.,  2018, \mn@doi [The
  Astrophysical Journal] {10.3847/1538-4357/aaae00}, 856, 7

\bibitem[\protect\citeauthoryear{Jaeggli \& Norton}{Jaeggli \&
  Norton}{2016}]{Jaeggli:2016}
Jaeggli S.~A.,  Norton A.~A.,  2016, \mn@doi [The Astrophysical Journal
  Letters] {10.3847/2041-8205/820/1/L11}, 820, 4pp

\bibitem[\protect\citeauthoryear{James, Witten, Hastie  \& Tibshirani}{James
  et~al.}{2013}]{James:2013}
James G.,  Witten D.,  Hastie T.,   Tibshirani R.,  2013, {An Introduction to
  Statistical Learning}, 1st edn.
Springer, New York, NY, USA

\bibitem[\protect\citeauthoryear{Jolliffe \& Stephenson}{Jolliffe \&
  Stephenson}{2003}]{Jolliffe:2003}
Jolliffe I.~T.,  Stephenson D.~B.,  2003, {Forecast Verification - A
  Practitioner's Guide in Atmospheric Science}, 1st edn.
John Wiley {\&} Sons, Chichester, England

\bibitem[\protect\citeauthoryear{Jonas, Bobra, Shankar, Hoeksema  \&
  Recht}{Jonas et~al.}{2018}]{Jonas:2018}
Jonas R.,  Bobra M.,  Shankar V.,  Hoeksema J.~T.,   Recht B.,  2018, \mn@doi
  [Solar Physics] {10.1007/s11207-018-1258-9}, 293, 48

\bibitem[\protect\citeauthoryear{Kubo, Den  \& Ishii}{Kubo
  et~al.}{2017}]{Kubo:2017}
Kubo Y.,  Den M.,   Ishii M.,  2017, \mn@doi [Journal of Space Weather and
  Space Climate] {10.1051/swsc/2017018}, 7, A20

\bibitem[\protect\citeauthoryear{Lan, Jiang, Ding  \& Yang}{Lan
  et~al.}{2012}]{Lan:2012}
Lan R.-S.,  Jiang Y.,  Ding L.-G.,   Yang J.-W.,  2012, \mn@doi [Research in
  Astronomy and Astrophysics] {10.1088/1674-4527/12/9/002}, 12, 1191

\bibitem[\protect\citeauthoryear{Lee, Moon, Kim, Park  \& Fletcher}{Lee
  et~al.}{2007}]{Lee:2007}
Lee J.-Y.,  Moon Y.-J.,  Kim K.-S.,  Park Y.-D.,   Fletcher A.-B.,  2007,
  \mn@doi [Journal of The Korean Astronomical Society]
  {10.5303/JKAS.2007.40.4.099}, 40, 99

\bibitem[\protect\citeauthoryear{Leka, Barnes  \& Wagner}{Leka
  et~al.}{2018}]{Leka:2018}
Leka K.~D.,  Barnes G.,   Wagner E.,  2018, \mn@doi [Journal of Space Weather
  and Space Climate] {10.1051/swsc/2018004}, 8, A25

\bibitem[\protect\citeauthoryear{Li \& Zhu}{Li \& Zhu}{2013}]{Li:2013}
Li R.,  Zhu J.,  2013, \mn@doi [Research in Astronomy and Astrophysics]
  {10.1088/1674-4527/13/9/010}, 13, 1118

\bibitem[\protect\citeauthoryear{Li, Cui, He  \& Wang}{Li
  et~al.}{2008}]{Li:2008}
Li R.,  Cui Y.,  He H.,   Wang H.,  2008, \mn@doi [Advances in Space Research]
  {10.1016/j.asr.2007.12.015}, 42, 1469

\bibitem[\protect\citeauthoryear{Li, Wang, Cui  \& Huang}{Li
  et~al.}{2011}]{Li:2011}
Li R.,  Wang H.,  Cui Y.,   Huang X.,  2011, \mn@doi [Science China Physics,
  Mechanics and Astronomy] {10.1007/s11433-011-4391-0}, 54, 1546

\bibitem[\protect\citeauthoryear{Liu, Li, Wan  \& Yu}{Liu
  et~al.}{2017a}]{Liu:2017a}
Liu J.-F.,  Li F.,  Wan J.,   Yu D.-R.,  2017a, \mn@doi [Research in Astronomy
  and Astrophysics] {10.1088/1674-4527/17/4/34}, 17, 034

\bibitem[\protect\citeauthoryear{Liu, Li, Zhang  \& Yu}{Liu
  et~al.}{2017b}]{Liu:2017b}
Liu J.-F.,  Li F.,  Zhang H.-P.,   Yu D.-R.,  2017b, \mn@doi [Research in
  Astronomy and Astrophysics] {10.1088/1674-4527/17/11/116}, 17, 116

\bibitem[\protect\citeauthoryear{Liu, Deng, Wang  \& Wang}{Liu
  et~al.}{2017c}]{Liu:2017c}
Liu C.,  Deng N.,  Wang J. T.~L.,   Wang H.,  2017c, \mn@doi [The Astrophysical
  Journal] {10.3847/1538-4357/aa789b}, 843, 104

\bibitem[\protect\citeauthoryear{Mason \& Hoeksema}{Mason \&
  Hoeksema}{2010}]{Mason:2010}
Mason J.~P.,  Hoeksema J.~T.,  2010, \mn@doi [The Astrophysical Journal]
  {10.1088/0004-637X/723/1/634}, 723, 634

\bibitem[\protect\citeauthoryear{McAteer, Gallagher  \& Conlon}{McAteer
  et~al.}{2010}]{McAteer:2010}
McAteer R. T.~J.,  Gallagher P.~T.,   Conlon P.~A.,  2010, \mn@doi [Advances in
  Space Research] {10.1016/j.asr.2009.08.026}, 45, 1067

\bibitem[\protect\citeauthoryear{McCloskey, Gallagher  \& Bloomfield}{McCloskey
  et~al.}{2018}]{McCloskey:2018}
McCloskey A.~E.,  Gallagher P.~T.,   Bloomfield D.~S.,  2018, \mn@doi [Journal
  of Space Weather and Space Climate] {10.1051/swsc/2018022}, 8, A34

\bibitem[\protect\citeauthoryear{McIntosh}{McIntosh}{1990}]{McIntosh:1990}
McIntosh P.~S.,  1990, \mn@doi [Solar Physics] {10.1007/BF00158405}, 125, 251

\bibitem[\protect\citeauthoryear{Messerotti, Zuccarello, Guglielmino, Bothmer,
  Lilensten, Noci, Storini  \& Lundstedt}{Messerotti
  et~al.}{2009}]{Messerotti:2009}
Messerotti M.,  Zuccarello F.,  Guglielmino S.~L.,  Bothmer V.,  Lilensten J.,
  Noci G.,  Storini M.,   Lundstedt H.,  2009, \mn@doi [Space Science Reviews]
  {10.1007/s11214-009-9574-x}, 147, 121

\bibitem[\protect\citeauthoryear{Miller}{Miller}{1988}]{Miller:1988}
Miller R.,  1988, The Royal Astronomical Society of Canada, 82, 191

\bibitem[\protect\citeauthoryear{Muranushi, Shibayama, Muranushi, Isobe,
  Nemoto, Komazaki  \& Shibata}{Muranushi et~al.}{2015}]{Muranushi:2015}
Muranushi T.,  Shibayama T.,  Muranushi Y.~H.,  Isobe H.,  Nemoto S.,  Komazaki
  K.,   Shibata K.,  2015, \mn@doi [Space Weather] {10.1002/2015SW001257}, 13,
  778

\bibitem[\protect\citeauthoryear{Murray, Bingham, Sharpe  \& Jackson}{Murray
  et~al.}{2017}]{Murray:2017}
Murray S.~A.,  Bingham S.,  Sharpe M.,   Jackson D.~R.,  2017, \mn@doi [Space
  Weather] {10.1002/2016SW001579}, 15, 577

\bibitem[\protect\citeauthoryear{NRC}{NRC}{2009}]{NRC:2009}
NRC N. R.~C.,  2009, in Committee on the Societal and Economic Impacts of
  Severe Space Weather Events: A Workshop. The National Academic Press,
  Washington, DC, pp 1--131

\bibitem[\protect\citeauthoryear{Nishizuka, Sugiura, Kubo, Den, Watari  \&
  Ishii}{Nishizuka et~al.}{2017}]{Nishizuka:2017}
Nishizuka N.,  Sugiura K.,  Kubo Y.,  Den M.,  Watari S.,   Ishii M.,  2017,
  \mn@doi [The Astrophysical Journal] {10.3847/1538-4357/835/2/156}, 835, 156
  (10pp)

\bibitem[\protect\citeauthoryear{Nishizuka, Sugiura, Kubo, Den  \&
  Ishii}{Nishizuka et~al.}{2018}]{Nishizuka:2018}
Nishizuka N.,  Sugiura K.,  Kubo Y.,  Den M.,   Ishii M.,  2018, \mn@doi [The
  Astrophysical Journal] {10.3847/1538-4357/aab9a7}, 858, 8pp

\bibitem[\protect\citeauthoryear{Park, Moon, Shin, Yi, Lim, Lee  \& Shin}{Park
  et~al.}{2018}]{Park:2018}
Park E.,  Moon Y.-J.,  Shin S.,  Yi K.,  Lim D.,  Lee H.,   Shin G.,  2018,
  \mn@doi [The Astrophysical Journal] {10.3847/1538-4357/aaed40}, 869, 6pp

\bibitem[\protect\citeauthoryear{Pyle}{Pyle}{1999}]{Pyle:1999}
Pyle D.,  1999, {Data Preparation for Data Mining}.
No. March 2012, Morgan Kaufmann Publishers, San Francisco, USA

\bibitem[\protect\citeauthoryear{Qahwaji \& Colak}{Qahwaji \&
  Colak}{2006}]{Qahwaji:2006}
Qahwaji R.,  Colak T.,  2006, in Proceedings of the 3rd International
  Conference on Cybernatics and Information Technologies. Florida, US

\bibitem[\protect\citeauthoryear{Qahwaji \& Colak}{Qahwaji \&
  Colak}{2007}]{Qahwaji:2007}
Qahwaji R.,  Colak T.,  2007, \mn@doi [Solar Physics]
  {10.1007/s11207-006-0272-5}, 241, 195

\bibitem[\protect\citeauthoryear{Raboonik, Safari, Alipour  \&
  Wheatland}{Raboonik et~al.}{2016}]{Raboonik:2016}
Raboonik A.,  Safari H.,  Alipour N.,   Wheatland M.~S.,  2016, \mn@doi [The
  Astrophysical Journal] {10.3847/1538-4357/834/1/11}, 834, 11 (8pp)

\bibitem[\protect\citeauthoryear{Sadykov \& Kosovichev}{Sadykov \&
  Kosovichev}{2017}]{Sadykov:2017b}
Sadykov V.~M.,  Kosovichev A.~G.,  2017, \mn@doi [The Astrophysical Journal]
  {10.3847/1538-4357/aa9119}, 849, 148

\bibitem[\protect\citeauthoryear{Sarkar, Bali  \& Sharma}{Sarkar
  et~al.}{2018}]{Sarkar:2018}
Sarkar D.,  Bali R.,   Sharma T.,  2018, {Practical Machine Learning with
  Python}, 1st edn.
Springer Science + Business Media, New York

\bibitem[\protect\citeauthoryear{Scherrer et~al.,}{Scherrer
  et~al.}{1995}]{Scherrer1995a}
Scherrer P.~H.,  et~al., 1995, \mn@doi [Solar Physics] {10.1007/BF00733429},
  162, 129

\bibitem[\protect\citeauthoryear{Shin, Lee, Moon, Chu  \& Park}{Shin
  et~al.}{2016}]{Shin:2016}
Shin S.,  Lee J.-Y.,  Moon Y.-J.,  Chu H.,   Park J.,  2016, \mn@doi [Solar
  Physics] {10.1007/s11207-016-0869-2}, 291, 897

\bibitem[\protect\citeauthoryear{Song, Tan, Jing, Wang, Yurchyshyn  \&
  Abramenko}{Song et~al.}{2009}]{Song:2009}
Song H.,  Tan C.,  Jing J.,  Wang H.,  Yurchyshyn V.,   Abramenko V.,  2009,
  \mn@doi [Solar Physics] {10.1007/s11207-008-9288-3}, 254, 101

\bibitem[\protect\citeauthoryear{Wang, Cui, Li, Zhang  \& Han}{Wang
  et~al.}{2008}]{Wang:2008}
Wang H.,  Cui Y.,  Li R.,  Zhang L.,   Han H.,  2008, \mn@doi [Advances in
  Space Research] {10.1016/j.asr.2007.06.070}, 42, 1464

\bibitem[\protect\citeauthoryear{Wheatland}{Wheatland}{2005}]{Wheatland:2005}
Wheatland M.~S.,  2005, \mn@doi [Space Weather] {10.1029/2004SW000131}, 3

\bibitem[\protect\citeauthoryear{Wilson}{Wilson}{1972}]{Wilson:1972}
Wilson D.~L.,  1972, IEEE Transactions on Systems, Man and Cybernetics, 2, 408

\bibitem[\protect\citeauthoryear{Winter \& Balasubramaniam}{Winter \&
  Balasubramaniam}{2015}]{Winter:2015}
Winter L.~M.,  Balasubramaniam K.,  2015, \mn@doi [Space Weather]
  {10.1002/2015SW001170}, 13, 286

\bibitem[\protect\citeauthoryear{Witten, Frank  \& Hall}{Witten
  et~al.}{2011}]{Witten:2011}
Witten I.~H.,  Frank E.,   Hall M.~A.,  2011, {Data Mining: Practical Machine
  Learning Tools and Techniques}, 3 edn.
Morgan Kaufmann Publishers, Burlington

\bibitem[\protect\citeauthoryear{Yang, Lin, Zhang  \& Mao}{Yang
  et~al.}{2013}]{Yang:2013}
Yang X.,  Lin G.,  Zhang H.,   Mao X.,  2013, \mn@doi [The Astrophysical
  Journal] {10.1088/2041-8205/774/2/L27}, 774, 6pp

\bibitem[\protect\citeauthoryear{Yu, Huang, Wang  \& Cui}{Yu
  et~al.}{2009}]{Yu:2009}
Yu D.,  Huang X.,  Wang H.,   Cui Y.,  2009, \mn@doi [Solar Physics]
  {10.1007/s11207-009-9318-9}, 255, 91

\bibitem[\protect\citeauthoryear{Yu, Huang, Hu, Zhou, Wang  \& Cui}{Yu
  et~al.}{2010a}]{Yu:2010a}
Yu D.,  Huang X.,  Hu Q.,  Zhou R.,  Wang H.,   Cui Y.,  2010a, \mn@doi [The
  Astrophysical Journal] {10.1088/0004-637X/709/1/321}, 709, 321

\bibitem[\protect\citeauthoryear{Yu, Huang, Wang, Cui, Hu  \& Zhou}{Yu
  et~al.}{2010b}]{Yu:2010b}
Yu D.,  Huang X.,  Wang H.,  Cui Y.,  Hu Q.,   Zhou R.,  2010b, \mn@doi [The
  Astrophysical Journal] {10.1088/0004-637X/710/1/869}, 710, 869

\bibitem[\protect\citeauthoryear{Yuan, Shih, Jing  \& Wang}{Yuan
  et~al.}{2010}]{Yuan:2010}
Yuan Y.,  Shih F.~Y.,  Jing J.,   Wang H.-M.,  2010, \mn@doi [Research in
  Astronomy and Astrophysics] {10.1088/1674-4527/10/8/008}, 10, 785

\bibitem[\protect\citeauthoryear{Zaki \& Junior}{Zaki \&
  Junior}{2013}]{Zaki:2013}
Zaki M.~J.,  Junior W.~M.,  2013, {Data Mining and Analysis: Fundamental
  Concepts and Algorithms}.
Cambridge University Press, New York

\bibitem[\protect\citeauthoryear{Zhang, Liu  \& Wang}{Zhang
  et~al.}{2011}]{Zhang:2011}
Zhang X.,  Liu J.,   Wang Q.,  2011, in 4th International Congress on Image and
  Signal Processing. pp 910--914

\bibitem[\protect\citeauthoryear{Zhang, Bi  \& Soda}{Zhang
  et~al.}{2017}]{Zhang:2017}
Zhang C.,  Bi J.,   Soda P.,  2017, in Proceedings of the 2017 IEEE
  International Conference on Bioinformatics and Biomedicine, BIBM 2017. IEEE,
  Kansas, USA, pp 933--938, \mn@doi{10.1109/BIBM.2017.8217782}

\makeatother
\end{thebibliography}








\bsp	
\label{lastpage}

\end{document}